\documentclass[prb,aip,reprint,twocolumn]{revtex4-1}
\usepackage{lipsum}% http://ctan.org/pkg/lipsum
\usepackage{graphicx} 
\usepackage{bm} 
\usepackage{dcolumn}
\usepackage{amssymb,amsmath}
\bibstyle{apsrev}
\usepackage{hyperref}
\usepackage{color}
\usepackage[normalem]{ulem}

\begin{document}

\title{Majorana Bound States hallmark in a quantum topological interferometer ring }%\sout{Using shot noise to detect Majorana Bound States}}
\author{A. M. Calle} 
\affiliation{Departamento de F\'isica, Universidad T\'ecnica Federico Santa Mar\'ia, Casilla Valpara\'iso, Chile}       
\author{P. A. Orellana} 
\affiliation{Departamento de F\'isica, Universidad T\'ecnica Federico Santa Mar\'ia, Casilla Valpara\'iso, Chile}        
\author{J. A. Ot\'alora}
\affiliation{Departamento de F\'{\i }sica, Universidad Cat\'{o}lica
del Norte, Casilla 1280, Antofagasta, Chile}

\begin{abstract}
In this work, we investigate the conductance and current correlations properties of a quantum topological inteferometer consisting of a QD coupled to two Majorana Bound States (MBSs) confined at both ends of a 1D topological superconductor ring nanowire. We analyze the ring in its topological non trivial and trivial phases to show that the tunneling conductance, shot noise and fano factor 
%have unique behavior in the former case, 
present unique characteristics to distinguish the hallmark of MBSs. We reinforce our findings by taking advantage of the correspondence between the quantum topological interferometer and a dot effectively coupled to a single Majorana state in a straight topological superconductor wire configuration. We show that, 
besides the characteristic zero-bias conductance $e^{2} /2h$ and the already known shot noise features, the Fano factor provides significant  information to distinguish the MBSs presence.

\end{abstract}
\maketitle

\section{Introduction}

A Majorana bound state (MBS), in condensed matter physics, is a zero-energy quasi-particle with the particularity of being its own antiparticle. \cite{majorana1937teoria} These quasi-particles, belongs to the family of anyons and therefore have a non-Abelian exchange statistics which makes them very interesting objects for fault tolerant topological quantum computation. \cite{Alicea, Beenakker2,Yoreg, DasSarma, Hasan, Sato}

%The type of semi- conductor is important: it must have a strong spin-orbit coupling and a large Zeeman effect (spectral line splitting in a magnetic field). The spin-orbit coupling locks the spin orientation of the charge carriers with respect to their momentum direction, which affects the way the electrons pair up in the superconductor. The end result is a switch from s-wave superconductivityÑin which the superconducting energy gap is isotropicÑto p-wave superconductivityÑin which the gap depends on the momentum direction. Above a critical magnetic field, the p-wave superconductor in the nanowires can become topologically nontrivial,

MBSs take place in quantum systems with strong spin-orbit coupling, superconductivity, and broken time-reversal symmetry. \cite{Yoreg, DasSarma, Fulga}
%Currently, the most promising experimental setup for observing MBSs consists of a semiconductor nanowire in proximity to a superconductor.
The most promising platforms to observe MBSs involve topological superconductors realized in semiconductors, specifically, semiconductor nanowires with a strong spin orbit coupling in proximity to an s-wave superconductor and subject to a magnetic field.\cite{Yoreg, DasSarma, Fulga,DasSarma2011, Cayao}
The spin-orbit coupling affects dramatically the way the electrons pair up in the superconductor, resulting in a switch from s-wave superconductivity to p-wave superconductivity, along with the magnetic field which will drive the p-wave superconductor to a topological phase transition.
%Above a critical magnetic field, the p-wave superconductor in the nanowires can become topologically nontrivial. 
Theory predicts that the boundaries of this topological superconductor --- in this case the ends of the nanowire --- should host MBSs. \cite{Phys13}.

 % The main advantage of the semiconductor/superconductor proposal is its simplicity: it includes a conventional semiconductor with strong Rashba coupling such as InAs or InSb, a conventional superconductor such as Al or Nb, and an in-plane magnetic field. \cite{DasSarma2011, Cayao} In essence, the 
A criteria to detect Majorana modes consist on measuring the zero-bias conductance peak (ZBCP) from tunneling electrons into the MBSs. \cite{Mourik, Das, Finck, Rokhinson, Nichele, Flensberg}  Nevertheless, confirming such states require seeking for extra features since other zero energy modes different than MBSs can also lead to zero bias peaks, for instance, from Andreev bound states (ABSs), multi-band effects \cite{Multiband}, weak antilocalization \cite{weak_anti} and the Kondo effect. \cite{goldhaber1998kondo, Hell} 
Currently, distinguishing MBSs from ABSs is one of the most critical challenges, which has lead to considerable theoretical proposals,\cite{Fingerprints, Yoreg1, Hell, DasSarma1, DasSarma2, Seridonio, Deng2018, Sau} mostly focused on quantum dots coupled to topological superconductors (QD--MBSs configurations). Indeed, evidence of their existence have been shown by probing their transport conductance spectrum,\cite{PhysRevB.84.201308, Cao2012,Seridonio} thermal conductance \cite{ricco2018tuning,Leijnse_2014}, ac Josephson effect \cite{PhysRevB.86.140503, PhysRevLett.108.257001} and current noise correlations.\cite{Lu2016, Cao2012,Yoreg1,Chen_2014,PhysRevB.96.115413,PhysRevLett.122.097003,PhysRevB.91.081405,PhysRevB.102.045303} Particularly, it has been proposed to combine tunneling conductance and shot noise correlations measurements as a complementary diagnosis method to distinguish real from fake MBSs.\cite{Chen_2014, PhysRevLett.124.096801, guerci2019probing, PhysRevX.4.031051}

In this letter, besides of studying the tunneling conductance and shot noise correlations properties, we also focus on seeking a  distinguishable fano factor fingerprint of a QD coupled to two MBSs confined at the ends of a 1D topological superconductor nanowire ring --- denoted here as QD--MBSs ring system. We underpin our findings by analyzing the conditions that lead to a full correspondence between our system of interest (QD--MBSs ring) and a topological QD--MBSs wire system,\cite{PhysRevB.84.201308} as illustrated in Figure \ref{system}. Finally, we argue that the fano factor jointly with the reported results for ZBCP and shot noise, would stand as a more robust diagnosis tool for distinguishing the real MBSs from spurious-zero energy modes.

%In this paper, we discuss the signatures of MBSs in spin-resolved current correlations
%an MBS signature can be extracted from this ÒnoiseÓ
%the Fano factor exhibits different behaviors depending on whether the nanowires contain true MBSs or ABS impostors.
%As we will show, in the presence of a MBS, the cross term P"# carries unique signatures, that are strikingly di?erent from the case of an ABS: 
%We shall argue that measuring the Fano factor would provide a clear signature of the Majorana nature of a bound state.

\section{Description of the Model}

\begin{figure}[t]
\centerline{\includegraphics[scale=0.3,angle=0]{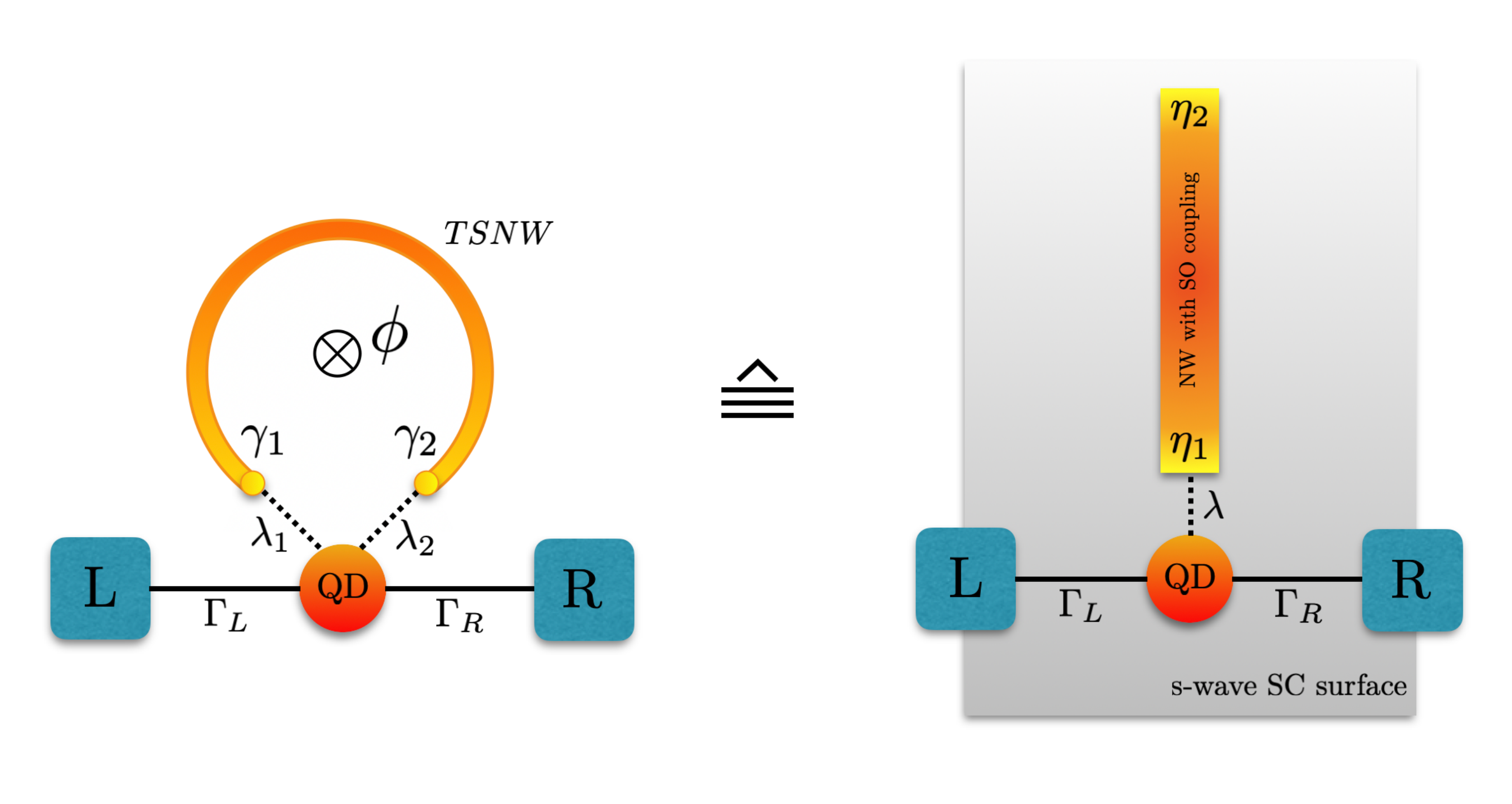}}
\caption{Schematic setup of the QD-MBSs ring system. A QD is coupled to two MBSs, $\gamma_{1}$ and $\gamma_{2}$, located at the ends of a TSNW. Here, $\lambda_{1}=|\lambda_{1}|e^{i\phi/4}$, $\lambda_{2}=|\lambda_{2}|e^{-i\phi/4}$, where  $|\lambda_{1}|$ and $|\lambda_{2}|$ denote the QD-MBS coupling strength and $\phi=\Phi/\Phi_{0}$ with $\Phi_{0}=h/2e$ is the phase factor resulting from the threading magnetic flux. Two normal metallic leads L and R are attached to the QD with coupling strength $\Gamma_{L}$ and $\Gamma_{R}$. As we will show the topological QD--MBSs ring system (a) is equivalent to the topological QD--MBSs wire system configuration in (b).}
\label{system}
\end{figure}

We consider the setup shown in Fig. \ref{system} (a) in which a spinless quantum dot is coupled to two MBSs, $\gamma_{1}$ and $\gamma_{2}$, located at the ends of a TSNW. \cite{PhysRevB.97.035310}
The Hamiltonian takes the form
\begin{equation}
\label{H}
H=H_{Leads}+H_{Dot}+H_{MBS}+H_{DM}+H_{T}
\end{equation}
\noindent Where, $H_{Leads}$ describes the left (L) and right (R) metallic leads, 
\begin{equation}
H_{Leads}=\sum_{k,\alpha=L,R}\epsilon_{k\alpha}c^{\dag}_{k\alpha}c_{k\alpha}
\end{equation}
\noindent $c^{\dag}_{k\alpha}$ and $c_{k\alpha}$ are the creation and annihilation operators with energy $\epsilon_{k\alpha}$ in the lead $\alpha=L, R$.
$H_{Dot}$ is the Hamiltonian of the quantum dot,  
\begin{equation}
H_{Dot}=\epsilon_{d}d^{\dag}d
\end{equation}
\noindent which describes a dot with an energy level $\epsilon_{d}$, with $d^{\dag}$ ($d$) being its creation (annihilation) operator.
The term $H_{MBS}$ in Eq.(\ref{H})
\begin{equation}
H_{MBS}=i\epsilon_{M}\gamma_{1}\gamma_{2}
\end{equation}
\noindent describes the coupling between the two MBSs, $\gamma_{1}$ and $\gamma_{2}$, with the overlap being $\epsilon_{M}$.
The term $H_{DM}$ denotes the coupling between the QD and the MBSs
\begin{equation}
H_{DM}=\left(\lambda^{*}_{1}d^{\dag}-\lambda_{1}d\right)\gamma_{1}+i\left(\lambda^{*}_{2}d^{\dag}+\lambda_{2}d\right)\gamma_{2}
\label{HDM}
\end{equation}
\noindent with the coupling parameters $\lambda_{1}=|\lambda_{1}|e^{i\phi/4}$, $\lambda_{2}=|\lambda_{2}|e^{-i\phi/4}$, where $|\lambda_{1}|$ and $|\lambda_{2}|$ denote the respective coupling strength and $\phi=\Phi/\Phi_{0}$ with $\Phi_{0}=h/2e$ is the phase factor resulting from the threading magnetic flux. 
The last term in Eq.(\ref{H})
\begin{equation}
H_{T}=\sum_{k\alpha}\left(t_{\alpha}c^{\dag}_{k\alpha}d+h.c\right)
\end{equation}
\noindent describes the tunneling coupling between the QD and the lead $\alpha$ with strength $t_{\alpha}$.

The two MBSs $\gamma_{1}$ and $\gamma_{2}$ can be represented by their equivalent Dirac fermion operators according to $\gamma_{1}=\left(f^{\dag}+f\right)/\sqrt{2}$ and $\gamma_{2}=i\left(f^{\dag}-f\right)/\sqrt{2}$, which transforms the terms $H_{MBS}$ and $H_{DM}$ in the Hamiltonian as follows,
\begin{equation}
H_{MBS}=\epsilon_{M}\left(f^{\dag}f-\frac{1}{2}\right)
\end{equation}
\begin{eqnarray}
H_{DM}&=&\frac{1}{\sqrt{2}}\left(\lambda^{*}_{1}-\lambda^{*}_{2}\right)d^{\dag}f^{\dag}+\frac{1}{\sqrt{2}}\left(\lambda_{1}-\lambda_{2}\right)fd \\ \nonumber
&+&\frac{1}{\sqrt{2}}\left(\lambda^{*}_{1}+\lambda^{*}_{2}\right)d^{\dag}f+\frac{1}{\sqrt{2}}\left(\lambda_{1}+\lambda_{2}\right)f^{\dag}d
\end{eqnarray}

%In general, the current from the lead $\alpha$ ($\alpha$=L to $\alpha$=R) is given by $I_{\alpha}=e\langle \dot{N}_{\alpha} \rangle=\dot{\imath}\frac{e}{\hbar}\langle \left[H,N_{\alpha}\right] \rangle$, from which, we can get \cite{Jauho_book}

%\begin{equation}
%I_{L}=-2\frac{e}{h}\int d\epsilon \left[f_{L}\left(\epsilon\right)-f_{R}\left(\epsilon\right)\right]\frac{\Gamma_{L}\Gamma_{R}}{\Gamma_{L}+\Gamma_{R}}\Im\left[G^{r}_{1,11}\left(\epsilon\right)\right]\hspace{0.1cm}.
%\end{equation}
%\noindent Then, the zero-temperature conductance is
%\begin{equation}
%\label{condG}
%G=-\frac{2e^{2}}{h}\frac{\Gamma_{L}\Gamma_{R}}{\Gamma_{L}+\Gamma_{R}}\Im\left[G^{r}_{1,11}\left(\epsilon\right)\right]|_{\epsilon=eV}\hspace{0.1cm}.
%\end{equation}
In general, the current from the lead $\alpha$ ($\alpha$=L to $\alpha$=R) is given by $I_{\alpha}=e\langle \dot{N}_{\alpha} \rangle=\dot{\imath}\frac{e}{\hbar}\langle \left[H,N_{\alpha}\right] \rangle$, from which, we can get \cite{Jauho_book},
\begin{equation}
\label{Current}
\widehat{I}_{\alpha}\left(t\right)=\frac{i e}{\hbar}\sum_{k}\left[t_{\alpha}\langle c^{\dag}_{k\alpha}\left(t\right)d\left(t\right)\rangle-t^{*}_{\alpha}\langle d^{\dag}(t)c_{k\alpha}(t)\rangle\right]
\end{equation}

We are concerned with fuctuations of the current away from their average value. We thus introduce the operators $\delta \widehat{I}_{\alpha}\left(t\right)=\widehat{I}_{\alpha}\left(t\right)-\langle I_{\alpha}\left(t\right)\rangle$ and define
the spectral density of shot noise by the Fourier transformation of the current correlation \cite{blanter2000shot_footnote}
\begin{equation}
\label{Current_correlation}
\Pi_{\alpha\alpha'}\left(t,t'\right)=\langle\delta \widehat{I}_{\alpha}\left(t\right)\delta \widehat{I}_{\alpha'}\left(t'\right)\rangle+\langle\delta \widehat{I}_{\alpha'}\left(t'\right)\delta \widehat{I}_{\alpha}\left(t\right)\rangle
\end{equation}

Substituting the current operator Eq.(\ref{Current}) into the current correlation Eq.(\ref{Current_correlation}) and using the Wick's theorem, the correlation function can be expressed by the Green functions of the system. Then, applying the Fourier transformation over the times $t$ and $t'$, and using the relation $S_{\alpha\alpha'}\left(\Omega\right)\delta\left(\Omega+\Omega'\right)=\frac{1}{2\pi}\Pi_{\alpha\alpha'}\left(\Omega,\Omega'\right)$, we obtain the shot noise of self-correlation $S=S_{LL}\left(0\right)$ in the left terminal as
\begin{eqnarray}
\label{ShotN}
S=&-&\frac{2e^{2}}{h}\int d\epsilon\left[G^{r}_{d}(\epsilon)\Sigma^{<}_{L}(\epsilon)G^{r}_{d}(\epsilon)\Sigma^{>}_{L}(\epsilon) \right. \\  \nonumber
&+& \left. G^{r}_{d}(\epsilon)\Sigma^{<}_{L}(\epsilon)G^{>}_{d}(\epsilon)\left(\Sigma^{a}_{L}(\epsilon)-\Sigma^{r}_{L}(\epsilon)\right) \right. \\ \nonumber
&+&\left. G^{<}_{d}(\epsilon)\left(\Sigma^{r}_{L}(\epsilon)-\Sigma^{a}_{L}(\epsilon)\right)G^{>}_{d}(\epsilon)\Sigma^{r}_{L}(\epsilon) \right. \\ \nonumber
&+&\left. \left(\Sigma^{r}_{L}(\epsilon)-\Sigma^{a}_{L}(\epsilon)\right)G^{<}_{d}(\epsilon)\Sigma^{>}_{L}(\epsilon)G^{a}_{d}(\epsilon) \right. \\ \nonumber
&+&\left. G^{<}_{d}(\epsilon)\left(\Sigma^{a}_{L}(\epsilon)-\Sigma^{r}_{L}(\epsilon)\right)G^{r}_{d}(\epsilon)\Sigma^{>}_{L}(\epsilon) \right. \\ \nonumber
&+&\left. G^{>}_{d}(\epsilon)\Sigma^{<}_{L}(\epsilon)G^{a}_{d}(\epsilon)\left(\Sigma^{r}_{L}(\epsilon)-\Sigma^{a}_{L}(\epsilon)\right) \right. \\ \nonumber
&+&\left. G^{<}_{d}(\epsilon)\left(\Sigma^{a}_{L}(\epsilon)-\Sigma^{r}_{L}(\epsilon)\right)G^{>}_{d}(\epsilon)\Sigma^{a}_{L}(\epsilon) \right. \\ \nonumber
&+&\left. G^{a}_{d}(\epsilon)\Sigma^{<}_{L}(\epsilon)G^{a}_{d}(\epsilon)\Sigma^{>}_{L}(\epsilon) \right. \\ \nonumber
&-&\left. \left(\Sigma^{<}_{L}(\epsilon)G^{>}_{d}(\epsilon)+\Sigma^{>}_{L}(\epsilon)G^{<}_{d}(\epsilon)\right)\right]
\end{eqnarray}

\noindent where $G^{r,a,<,>}_{d}(\epsilon)$ are the Green functions of the QD, $\Sigma^{r,a,<,>}_{L}$ are the self-energies of the L lead and $\Sigma^{r,a}_{L}=\mp i\Gamma_{L}/2$.
Where, $\Gamma_{L}=2\pi\rho_{L}V^{2}_{K_{L}}$ is the line width function describing the coupling between the dot and the $L$ lead in the wide band approximation, with $\rho_{L}$ being the density of states in the leads. 
%When $\Gamma=\Gamma_{L}+\Gamma_{R}$ is increased (decreased) the line width of the conductance increase (decrease). $\Im[G_{1,11}]$ denote the imaginary part of  the Green function $G_{1,11}$. $f_{L(R)}\left(\epsilon\right)=f\left(\epsilon-\mu_{L(R)}\right)$ is the Fermi-Dirac distribution with $\mu_{L(R)}$ the chemical potential for the lead $L(R)$

After some mathematical calculations we found the retarded Green function of the QD as follows \cite{Zeng2017}
\begin{equation}
\label{G_Zeng}
G^{r}_{d}\left(\omega\right)=\left[\omega-\epsilon_{d}+i\frac{\Gamma}{2}-A\left(\omega\right)-B\left(\omega\right)\right]^{-1}\hspace{0.1cm},
\end{equation}

\noindent where $A\left(\omega\right)=K\left(|\lambda_{1}|^{2}+|\lambda_{2}|^{2}+\frac{2\epsilon_{M}}{\omega}|\lambda_{1}||\lambda_{2}|\cos\frac{\phi}{2}\right)$ and $B\left(\omega\right)=\frac{K^{2}\left(|\lambda_{1}|^{4}+|\lambda_{2}|^{4}-2|\lambda_{1}|^{2}|\lambda_{2}|^{2}\cos\phi\right)}{\left(\omega+\epsilon_{d}+i\frac{\Gamma}{2}-A\left(\omega\right)\right)}$, with $K$ and $\Gamma$ being defined as $K=\frac{\omega}{\omega^{2}-\epsilon^{2}_{M}}$ and $\Gamma=\Gamma_{L}+\Gamma_{R}$.

Substituting all Green functions and self-energies into equation (\ref{ShotN})

\begin{eqnarray}
\label{sn_final}
S&=&\frac{2e^{2}}{h}\int d\epsilon \left[2|\lambda|^{4}|\widetilde{K}(\epsilon)|^{2}\Gamma^{2}_{L}|G^{r}_{d}(\epsilon)|^{2} F_{LL}(\epsilon)\right. \\ \nonumber
&+&\left.(1+C(\epsilon))^{2}T^{2}_{N}(\epsilon)\left[F_{LL}(\epsilon)+F_{RR}(\epsilon)\right] \right. \\ \nonumber
&+& \left. (1+C(\epsilon))T_{N}(\epsilon) \right. \\ \nonumber
&\times& \left. \{1-\left(1+C(\epsilon)\right)T_{N}(\epsilon)\} \left[F_{LR}(\epsilon)+F_{RL}(\epsilon)\right]\right.\Big]
\end{eqnarray}

\noindent where, 
\begin{equation}
C(\epsilon)=|\widetilde{K}(\epsilon)|^{2}\left(|\lambda_{1}|^{4}+|\lambda_{2}|^{4}-2|\lambda_{1}|^{2}|\lambda_{2}|^{2}\cos\phi\right)
\end{equation}

\noindent and $T_{N}=\Gamma_{L}\Gamma_{R}|G^{r}(\epsilon)|^{2}$ is the transmission. We also define $F_{\alpha\beta}(\epsilon)=f_{\alpha}(\epsilon)\left[1-f_{\beta}(\epsilon)\right]$, with $\alpha$ and $\beta$ being L and R. $f_{L(R)}\left(\epsilon\right)=f\left(\epsilon-\mu_{L(R)}\right)$ is the Fermi-Dirac distribution with $\mu_{L(R)}$ the chemical potential for the lead $L(R)$.
The first and the second term in equation (\ref{sn_final}) represent the thermal noise which vanish at zero temperature. Finally, we define $T_{N}=|\lambda|^{4}|\widetilde{K}(\epsilon)|^{2}T_{N}(\epsilon)$. Then the shot noise in 
equation (\ref{sn_final})  can be written as,

\begin{eqnarray} 
\label{Sn}
S&=&\frac{2e^{2}}{h}\int d\epsilon \left[ T_{N}\left(\epsilon\right)\left(1-T_{N}\left(\epsilon\right)\right)+T_{M}\left(\epsilon\right)\left(1-T_{M}\left(\epsilon\right)\right) \right.  \\ \nonumber
&-& \left. 2T_{N}\left(\epsilon\right)T_{M}\left(\epsilon\right)\right]\left(f_{L}\left(\epsilon\right)\left(1-f_{R}\left(\epsilon\right)\right)+f_{R}\left(\epsilon\right)\left(1-f_{L}\left(\epsilon\right)\right)\right)
\end{eqnarray}

\begin{figure*}[t!]
\centerline{\includegraphics[width=0.9\textwidth,scale=0.33,angle=0]{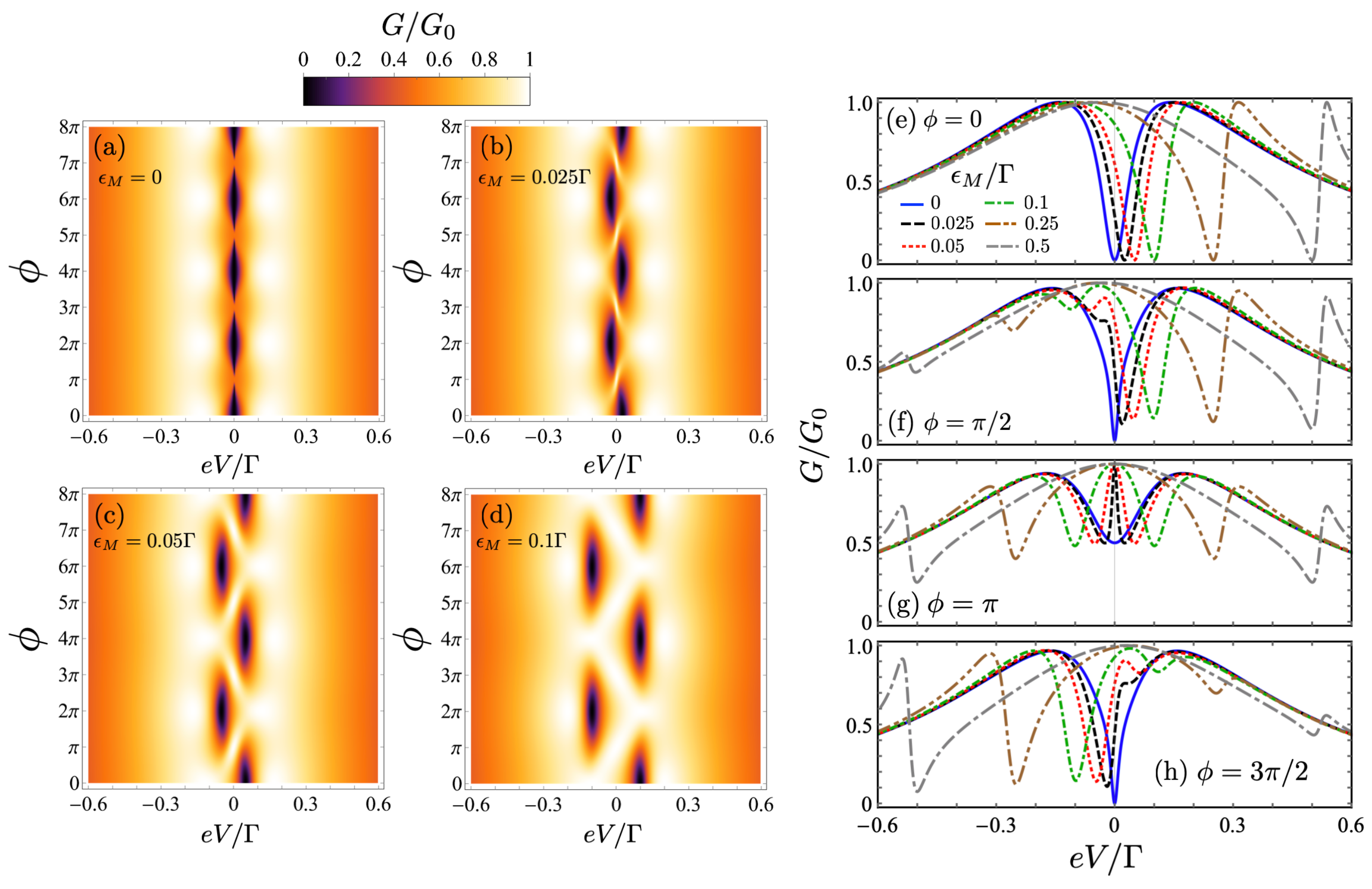}}
\caption{Differential conductance (in units of $G_{0}=e^{2}/h$) as a function of bias voltage $eV/\Gamma$ and $\phi$ for several values of MBSs coupling, $\epsilon_{M}$, (a) $\epsilon_{M}=0$, (b) $\epsilon_{M}=0.025\Gamma$, (c) $\epsilon_{M}=0.05\Gamma$, (d) $\epsilon_{M}=0.1\Gamma$. Figures (e) - (h) show the conductance as a function of bias voltage $eV/\Gamma$ for several values of $\epsilon_{M}$ and different values of magnetic flux phase $\phi$.
We use the following parameters: $|\lambda_{1}|=|\lambda_{2}|=0.1\Gamma$, $\epsilon_{d}=0$, $\Gamma_{L}=\Gamma_{R}=0.5\Gamma$.}
\label{conductance}
\end{figure*}

\begin{figure}[t]
\centerline{\includegraphics[scale=0.36,angle=0]{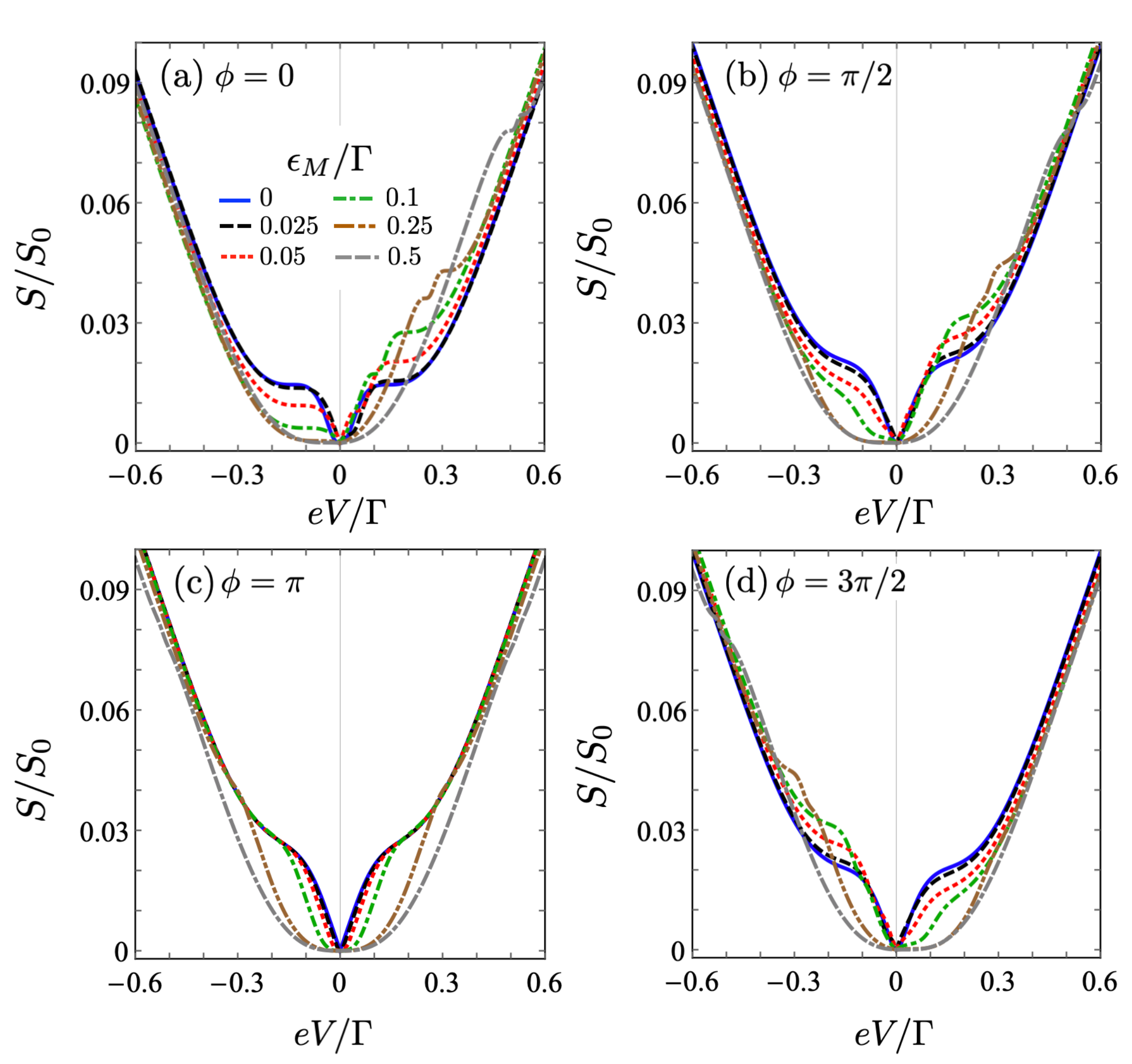}}
\caption{Shot noise (in units of $S_{0}=2e^{2}/h$) as a function of bias voltage $eV/\Gamma$ for several values of $\epsilon_{M}/\Gamma$ and for different values of magnetic flux phase, (a) $\phi=0$, (b) $\phi=\pi/2$, (c) $\phi=\pi$ and (d) $\phi=3\pi/2$. $|\lambda_{1}|=|\lambda_{2}|=0.1\Gamma$, $\epsilon_{d}=0$, $\Gamma_{L}=\Gamma_{R}=0.5\Gamma$}
\label{shot_noise}
\end{figure}

\section{Results}

In what follows, we set $\epsilon_{d}=0$ and assume that  the QD is symmetrically coupled to the two MBSs, that is, $|\lambda_{1}|=|\lambda_{2}|$. We also assume a symmetric dot-lead couplings $\Gamma_{L}=\Gamma_{R}$. From this point on, $\Gamma=\Gamma_{L}+\Gamma_{R}$ will be considered as the energy unit and $E_{F} = 0$. The shot noise is given in units of $S_{0} = 2e^{2}/h$.
%The shot noise and current are given in units of $S_{0} = 2e^{2}/h$ and $I_{0} = 2e/h$, respectively.

%In setups containing MBSs and QDs Fano resonances (FRs) for which a resonant path interferes with a continuous path [90] can emerge in the differential conductance.

In Figure \ref{conductance} (a)-(d), we show the conductance (in units of $G_{0}=e^{2}/h$) as a function of the bias voltage $eV/\Gamma$ and the magnetic flux phase $\phi$ for several values of coupling between MBSs, $\epsilon_{M}$. We can observe how the conductance changes periodically with the magnetic flux phase $\phi$, with the period being $2\pi$ when $\epsilon_{M}=0$ and  $4\pi$ for $\epsilon_{M}\neq0$ . The conductance as a function of $eV/\Gamma$ for several values of $\epsilon_{M}$ and for a few representatives values of $\phi$ is shown in Figure \ref{conductance} (e)-(f). In particular, when the magnetic flux is $2n\times2\pi$ or $(2n+1)\times2\pi$ only one Fano antiresonance (which emerge due to a resonant path interfering with a continuous path) whose minimal does fall to zero appear around $|eV|=\epsilon_{M}$ (see Fig. \ref{conductance} (e)). For the other values of $\phi$, two Fano antiresonances emerge approximately at $eV=\pm\epsilon_{M}$ whose minimum do not fall to zero (Fig. \ref{conductance} (f), (g))\cite{Zeng2017}. Especially when the nanowire is in its topological phase (the one with Majorana zero modes at the end of the nanowire), i.e., when $\phi=\pi, 3\pi,... \left(2n+1\right)\pi$, the antiresonances, located around $\pm\epsilon_{M}$ have an identical shape, but an opposite sign of the Fano parameter (Fig. \ref{conductance} (g)) \cite{AMCalle_Correspondence}. Regardless of the magnetic flux phase, as $\epsilon_{M}$ increases, the Fano antiresonance are shifted toward large values of $|eV|$. It is pertinent to mention here that the topological transition is associated with a substantial conductance variation. As is shown in Fig.  \ref{conductance} (a) for $\epsilon_{M}=0$, where we observe a jump from $G=0$ in the trivial topological region to $G = e^{2}/2h$ in the nontrivial topological region \cite{PhysRevB.84.201308}, which allows distinguishing the two different phases of the wire.

The results of shot noise calculated using Eq. (\ref{Sn})  are presented 
in Figure \ref{shot_noise}, where we show the shot noise (in units of $S_{0}=2e^{2}/h$) as a function of bias voltage $eV/\Gamma$ for several values of $\epsilon_{M}/\Gamma$ and different values of magnetic flux phase $\phi$. We notice that in analogy with the conductance (Fig. \ref{conductance}), when the magnetic flux changes, the shot noise changes periodically with a period $4\pi$.
When the coupling between Majorana fermions $\epsilon_{M}$ start to increase, we can observe how small steps appear in the shot noise. These small steps are positioned in the same value of $eV$ as the corresponding Fano antiresonances in the conductance. As $\epsilon_{M}$ increases, the height of these steps also increases, and they are shifted toward large $|eV|$. Particularly, it is interesting to notice that when the ring is in its topological phase, $\phi=\pi$,  (see Fig. \ref{shot_noise} (c)) these steps are not distinctly visible because the Fano antiresonances in the conductance do not fall to zero. Besides, the shot noise is symmetrical in the same way as the conductance.

\begin{figure*}[t]
\centerline{\includegraphics[width=0.9\textwidth,scale=0.33,angle=0]{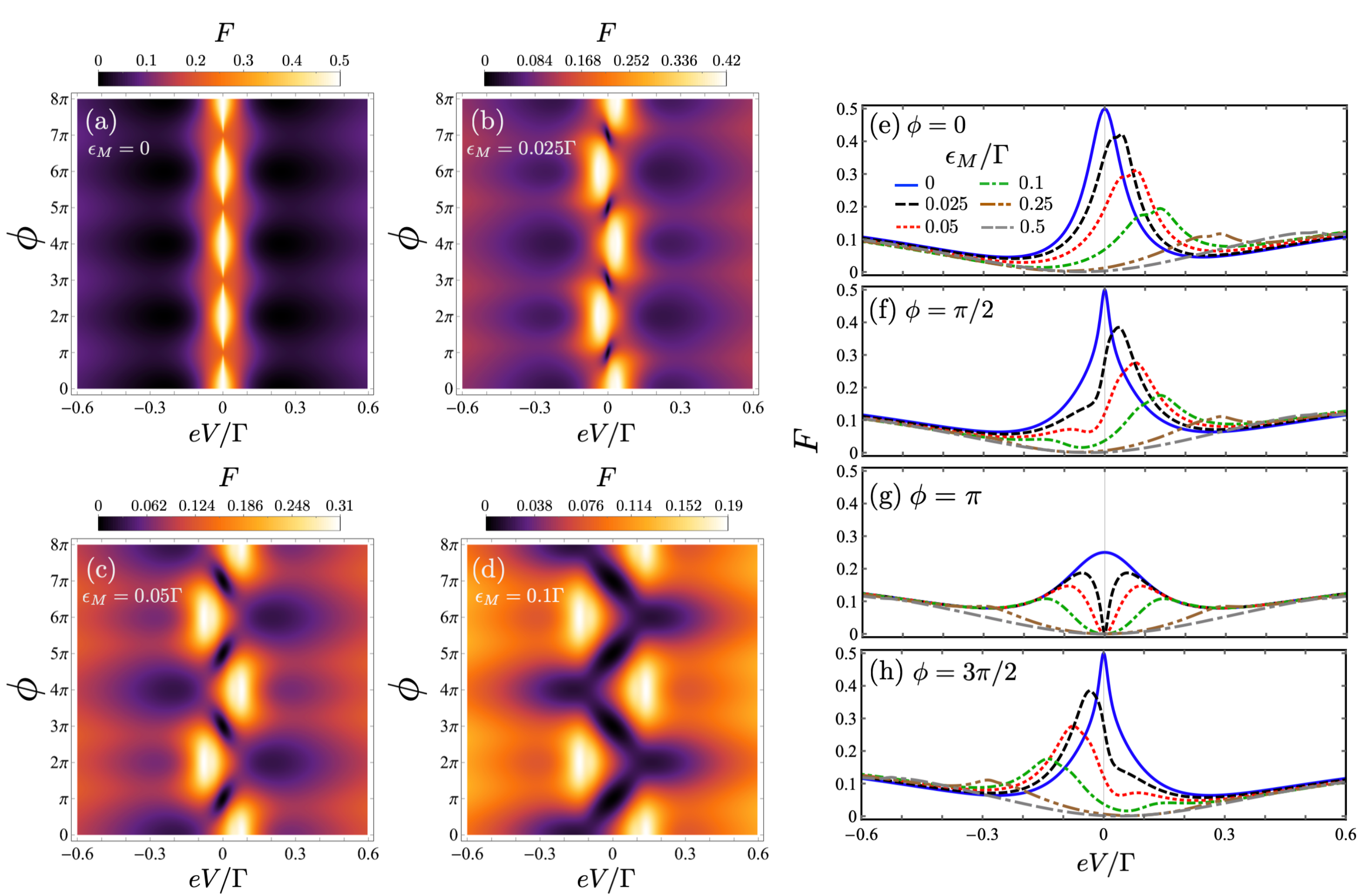}}
\caption{Fano factor as a function of bias voltage $eV/\Gamma$ and $\phi$ for several values of MBSs coupling, $\epsilon_{M}$, (a) $\epsilon_{M}=0$, (b) $\epsilon_{M}=0.025\Gamma$, (c) $\epsilon_{M}=0.05\Gamma$, (d) $\epsilon_{M}=0.1\Gamma$. Figures (e) - (h) show the Fano factor as a function of bias voltage $eV/\Gamma$ for several values of $\epsilon_{M}$ and different values of magnetic flux phase $\phi$.
We use the following parameters: $|\lambda_{1}|=|\lambda_{2}|=0.1\Gamma$, $\epsilon_{d}=0$, $\Gamma_{L}=\Gamma_{R}=0.5\Gamma$.}
\label{Fano_factor}
\end{figure*}

The calculation of shot noise and current allows to compute the Fano factor, defined as $F=S/2e|I|$, \cite{wan2005shot} which is shown
in Figure \ref{Fano_factor}. 
We display the evolution of the Fano factor as $\epsilon_{M}$ increase from $\epsilon_{M}=0$ up to $\epsilon_{M}=0.1\Gamma$ (see Figure \ref{Fano_factor} (a)-(d)). We observe that the Fano factor changes periodically as we sweep the magnetic flux phase $\phi$, which is a consequence of the periodicity in $\phi$ of the shot noise and current. We note that for $\epsilon_{M}=0$ the Fano factor is symmetrical, and we also observe how when we start to increase $\epsilon_{M}$ from $\epsilon_{M}=0.025\Gamma$  it becomes antisymmetric. It is relevant to notice that when $\epsilon_{M}=0$, there is a drastic variation of the Fano factor consisting in a jump from $F(V=0)=1/2$ in the trivial topological phase to $F(V=0)=1/4$ in the nontrivial topological phase of the system. This jump is related to the topological transition of the ring.
In Figures \ref{Fano_factor} (e)-(h) we can observe the Fano factor as a function of bias voltage $eV/\Gamma$ for several values of $\epsilon_{M}/\Gamma$ and for some representative values of magnetic flux phase $\phi$. We observe how when $\epsilon_{M}$ starts to increase, a tiny step located at the value of $|\epsilon_{M}|$ arise. The magnitude of this step decreases with $\epsilon_{M}$. Besides, we observe that when the ring is in its nontrivial topological phase ($\phi=\pi$), the Fano factor has different behavior compared with the Fano factor when the topological superconducting nanowire is in its trivial phase. When the ring is in its topological phase and $\epsilon_{M}=0$, the Fano factor acquires the value of $1/4$ at $eV=0$ (Fig.\ref{Fano_factor} (g)). When $\epsilon_{M}\neq0$ the value of the Fano factor at $eV=0$ is zero. This unique behavior does not occur when the ring is in the trivial topological phase, except for the case when there is a large overlap between the MBSs ($\epsilon_{M}$) at the two ends of the wire where for $\epsilon_{M}\neq0$ the Fano factor is zero at $eV=0$. This is caused by the fact that when $\epsilon_{M}$ is large, the two Majorana states are equivalent to a single ordinary ABS. \cite{Yoreg1}

In general, it is hard to find an analytical expression for the Fano factor. However, at zero bias voltage, the Fano factor can be approximated as follows: on the one hand, for $\epsilon_{M}=0$, we have $F=1/2$  and $F=\frac{1}{2}-\frac{2\Gamma_{L}\Gamma_{R}}{\Gamma^{2}}$, when the system in its trivial and nontrivial topological phases, respectively.  On the other hand, for  $\epsilon_{M}\neq0$, we obtain $F=\frac{1}{2}-\frac{2\Gamma_{L}\Gamma_{R}\epsilon^{2}_{M}}{(\Gamma^{2}+4\epsilon^{2}_{d})\epsilon^{2}_{M}+16\lambda^{2}_{1}\lambda^{2}_{2}\cos^{2}(\frac{\phi}{2})-16\lambda_{1}\lambda_{2}\cos(\frac{\phi}{2})\epsilon_{d}\epsilon_{M}}$.
%\begin{equation*}
%F=\frac{1}{2}-\frac{2\Gamma_{L}\Gamma_{R}\epsilon^{2}_{M}}{(\Gamma^{2}+4\epsilon^{2}_{d})\epsilon^{2}_{M}+16\lambda^{2}_{1}\lambda^{2}_{2}\cos^{2}(\frac{\phi}{2})-16\lambda_{1}\lambda_{2}\cos(\frac{\phi}{2})\epsilon_{d}\epsilon_{M}}
%\end{equation*}
As a consequence, when the system is in its topological phase, that is, $\phi=\left(2n+1\right)\pi$, the Fano factor is
$F=\frac{1}{2}-\frac{2\Gamma_{L}\Gamma_{R}}{\Gamma^{2}+4\epsilon^{2}_{d}}$.

The Fano factor exhibits different behavior depending on whether the nanowire is in its nontrivial or trivial topological phases. We will argue that an MBS signature can be extracted from this.
To this purpose, we note that the QD-Majorana coupling Hamiltonian,
$H_{Dot+MBS+DM}=H_{Dot}+H_{MBS}+H_{DM}$, for the MBS's -- QD system is, 
\begin{eqnarray}
H_{Dot+MBS+DM}&=&\epsilon_{d}d^{\dagger}d+i\epsilon_{M}\gamma_{1}\gamma_{2} \nonumber \\
&+&\left(\lambda^{*}_{1}d^{\dag}-\lambda_{1}d\right)\gamma_{1} \nonumber\\
&+&i\left(\lambda^{*}_{2}d^{\dag}+\lambda_{2}d\right)\gamma_{2}
\end{eqnarray}
%the Hamiltonian $H_{DM}$, which denotes the coupling between the QD and the two MBSs, $H_{DM}=\left(\lambda^{*}_{1}d^{\dag}-\lambda_{1}d\right)\gamma_{1}+i\left(\lambda^{*}_{2}d^{\dag}+\lambda_{2}d\right)\gamma_{2}$,
%as written in Equation (\ref{HDM})
%\begin{equation*}
%H_{DM}=\left(\lambda^{*}_{1}d^{\dag}-\lambda_{1}d\right)\gamma_{1}+i\left(\lambda^{*}_{2}d^{\dag}+\lambda_{2}d\right)\gamma_{2}
%\end{equation*}
which can be rewritten as the Hamiltonian of a dot that is effectively coupled to a single MBS. This can be made if we take (without lost of generality) $\lambda_{1}$ to be real ($\lambda_{1}=|\lambda_{1}|$) and $\lambda_{2}=|\lambda_{2}|e^{i\phi/2}$. 
Then, $H_{Dot+MBS+DM}$ reduces to 
\begin{eqnarray}
H_{Dot+MBS+DM}=\epsilon_{d}d^{\dagger}d&+&i\epsilon_{M}\left(\eta_{1}\eta_{2}-2i\frac{|\lambda_{1}||\lambda_{2}|}{\lambda^{2}}\cos(\phi/2)\right) \nonumber \\
&+&\lambda\left(\eta_{1}d^{\dag}-\eta^{\dag}_{1}d\right)
\end{eqnarray}
with $\eta_{1}=(|\lambda_{1}|\gamma_{1}+i|\lambda_{2}|e^{i\phi/2}\gamma_{2})/\lambda$, $\eta_{2}=(|\lambda_{1}|\gamma_{2}+i|\lambda_{2}|e^{-i\phi/2}\gamma_{1})/\lambda$
and $\lambda=\sqrt{|\lambda_{1}|^{2}+|\lambda_{2}|^{2}}$

The transformation can be written as:
\begin{equation}
\begin{pmatrix}
\eta_{1} \\
\eta_{2}
\end{pmatrix}=
\begin{pmatrix}
\cos\theta/2 & ie^{i\phi/2}\sin\theta/2 \\
ie^{-i\phi/2}\sin\theta/2 & \cos\theta/2
\end{pmatrix}
\begin{pmatrix}
\gamma_{1} \\
\gamma_{2}
\end{pmatrix}
\end{equation}

\noindent where, $\cos\theta/2=|\lambda_{1}|/\lambda$ and $\sin\theta/2=|\lambda_{2}|/\lambda$. This transformation belongs to the SU($2$) group.

\begin{figure}[t]
\centerline{\includegraphics[scale=0.4,angle=0]{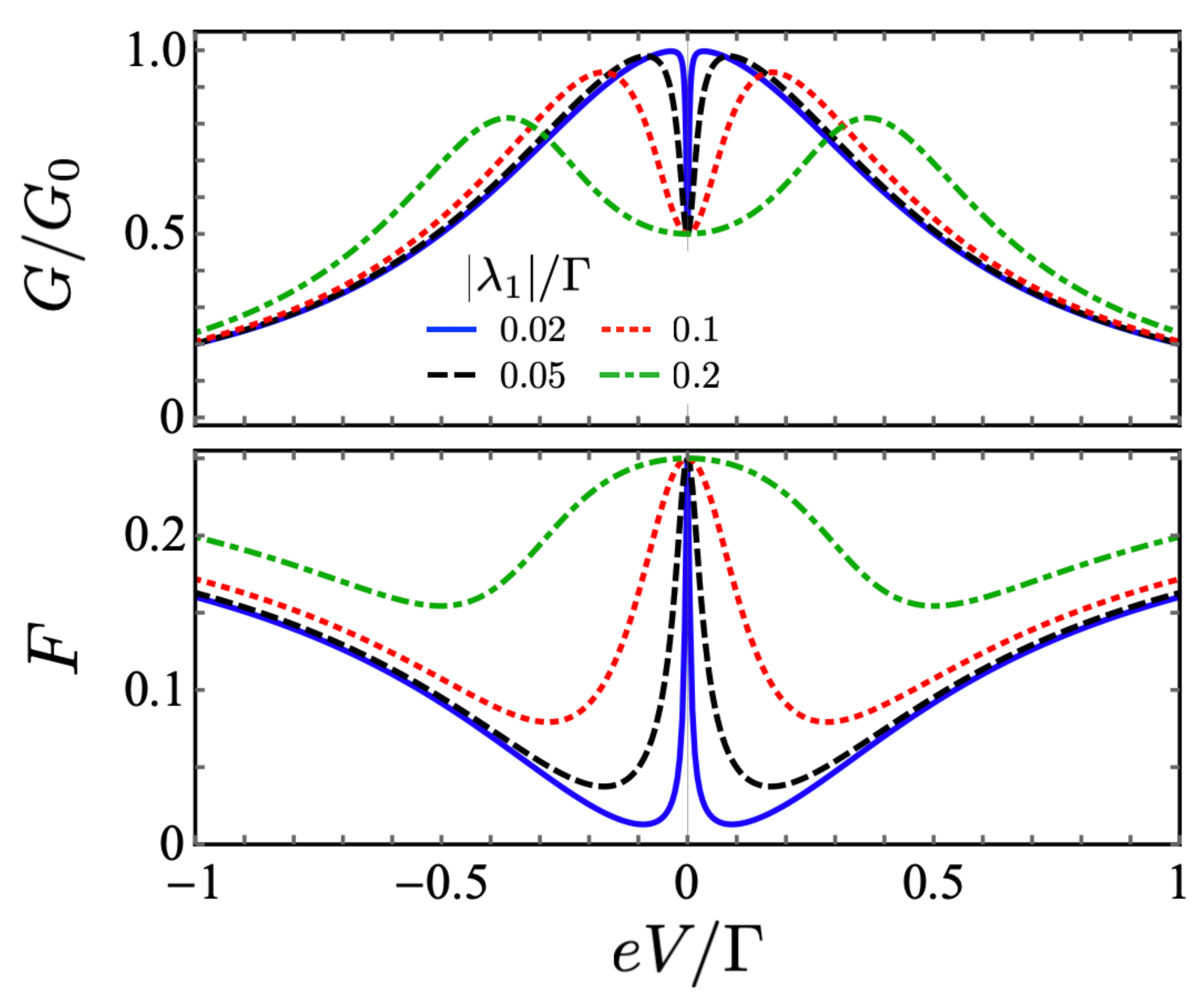}}
\caption{Conductance and Fano factor as a function of bias voltage $eV/\Gamma$ for several values of QD-MBSs coupling, $\lambda_{1}$ when $\epsilon_{M}=0$.
We use the following parameters: $\epsilon_{d}=0$, $\Gamma_{L}=\Gamma_{R}=0.5\Gamma$.}
\label{comparison}
\end{figure}

Note that when $\phi=\left(2n+1\right)\pi$ ($n$ integer), we obtain, $\eta_{1}=\eta^{\dag}_{1}$, $\eta_{2}=\eta^{\dag}_{2}$ and $H_{Dot+MBS+DM}=\epsilon_{d}d^{\dagger}d+i\epsilon_{M}\eta_{1}\eta_{2}+\lambda\left(d^{\dag}-d\right)\eta^{\dag}_{1}$, that is, a dot coupled to two MBSs reduces to a dot coupled to a single Majorana state $\eta_{1}$ which in turn is coupled to another Majorana fermion $\eta_{2}$ with a coupling $\epsilon_{M}$. \cite{PhysRevB.84.201308, PhysRevLett.106.090503} 
%the dot is effectively coupled to a single MBS 
Therefore, a QD coupled to two MBS in a ring configuration could be mapped into a dot effectively coupled to a single Majorana state, $\eta_{1}$, in a wire configuration for any value of $\epsilon_{M}$ (see Figure \ref{system}). It must be emphasized that this correspondence is independent of whether the magnitudes of the couplings ($|\lambda_{1}|$ and $|\lambda_{2}|$) are equal or not. As a consequence of such correspondence the conductance, shot noise and Fano factor for the QD-MBSs ring system and QD-MBS wire system are identical. 
In fact,  Figure \ref{comparison}, displays the differential conductance and Fano factor for different values of QD-MBSs coupling, $\lambda_{1}$, for both systems (as long as, $|\lambda_{1}|=|\lambda_{2}|$ in the QD--MBSs ring system and thus $|\lambda|=\sqrt{2}|\lambda_{1}|$ in the QD--MBS wire system).
As is well-known, the conductance in the topological non-trivial phase is always $G=e^{2}/2h$ \cite{PhysRevB.84.201308} as long as $\epsilon_{M} =0$, as can be seen in Figure \ref{comparison}.
It is worth noticing how the Fano factor increases with $\lambda_{1}$ for both configurations and more importantly, the Fano factor always gives $1/4$ at zero bias voltage, that is, $F(eV/\Gamma=0)=\frac{1}{4}$, as long as $\epsilon_{M}$=0, i.e. as long as the dot is coupled to a single MBS. 
Therefore, this result suggest that measurements of shot noise, in particular, of Fano factor give additional informations complementary to the one known by studying the characteristic zero-bias conductance $e^{2}/2h$.
%In general, it is hard to find an analytical expression for the Fano factor. However, at zero bias voltage, the Fano factor can be approximated as follows: on one hand, when $\epsilon_{M}=0$ we have $F=1/2$ when the system is not in its topological phase and $F=\frac{1}{2}-\frac{2\Gamma_{L}\Gamma_{R}}{\Gamma^{2}}$ for the system in its topological non-trivial phase. On the other hand, when $\epsilon_{M}\neq0$, we obtain 
%$F=\frac{1}{2}-\frac{2\Gamma_{L}\Gamma_{R}\epsilon^{2}_{M}}{(\Gamma^{2}+4\epsilon^{2}_{d})\epsilon^{2}_{M}+16\lambda^{2}_{1}\lambda^{2}_{2}\cos^{2}(\frac{\phi}{2})-16\lambda_{1}\lambda_{2}\cos(\frac{\phi}{2})\epsilon_{d}\epsilon_{M}}$ 
%\begin{equation*}
%F=\frac{1}{2}-\frac{2\Gamma_{L}\Gamma_{R}\epsilon^{2}_{M}}{(\Gamma^{2}+4\epsilon^{2}_{d})\epsilon^{2}_{M}+16\lambda^{2}_{1}\lambda^{2}_{2}\cos^{2}(\frac{\phi}{2})-16\lambda_{1}\lambda_{2}\cos(\frac{\phi}{2})\epsilon_{d}\epsilon_{M}}
%\end{equation*}
%As a consequence, when the system is in its topological phase ($\phi=\left(2n+1\right)\pi$) the Fano factor is
%$F=\frac{1}{2}-\frac{2\Gamma_{L}\Gamma_{R}}{\Gamma^{2}+4\epsilon^{2}_{d}}$.
In consequence, the study of the combination of both shot noise and conductance through a QD could provide a clear signature and allow to distinguish the MBSs. 
We believe that the predicted qualitative behavior of conductance and current correlations (shot noise)
could still hold when on-site Coulomb correlations are considered. \cite{Lutchyn_Probing,guerci2019probing} Besides, a study based on SBMF approach \cite{PhysRevX.4.031051} shows that a crossover from Kondo and Majorana dominated regimes can be realized by tuning the coupling $\lambda$. However, 
a detailed analysis of this problem is left to future investigations.

\section*{Summary}
In this work, we investigated the current correlations properties of a topological ring system configuration consisting of a QD coupled to two MBSs confined at both ends of a 1D topological superconductor nanowire. We found that when the ring is in its nontrivial topological phase, $\phi=\left(2n+1\right)\pi$, the Fano factor has a unique behavior compared with the Fano factor when the topological superconducting nanowire is in its trivial phase. 
To obtain an MBS distinguishing feature from this, we argued that a QD coupled to two MBS in a ring configuration could be mapped to a quantum-dot effectively connected to a single Majorana state in a wire configuration. As a consequence of such correspondence, we found that besides the characteristic zero-bias conductance $e^{2} /2h$, the Fano factor give additional information which could be definitive to find a clear signature to distinguish the MBSs.

\acknowledgments
This work was...

%%%%%%%%%%%%%%%%%%%%%%%%%%%%%%%%%%%%%%%%%%%%%%%%%%%%%%%%%%%%%%%%%%%%%
%% The appropriate \bibliographystyle and \bibliography commands
%% should be placed here.
%%%%%%%%%%%%%%%%%%%%%%%%%%%%%%%%%%%%%%%%%%%%%%%%%%%%%%%%%%%%%%%%%%%%%
\section*{References}
%\bibliography{References}

\begin{thebibliography}{51}%
\makeatletter
\providecommand \@ifxundefined [1]{%
 \@ifx{#1\undefined}
}%
\providecommand \@ifnum [1]{%
 \ifnum #1\expandafter \@firstoftwo
 \else \expandafter \@secondoftwo
 \fi
}%
\providecommand \@ifx [1]{%
 \ifx #1\expandafter \@firstoftwo
 \else \expandafter \@secondoftwo
 \fi
}%
\providecommand \natexlab [1]{#1}%
\providecommand \enquote  [1]{``#1''}%
\providecommand \bibnamefont  [1]{#1}%
\providecommand \bibfnamefont [1]{#1}%
\providecommand \citenamefont [1]{#1}%
\providecommand \href@noop [0]{\@secondoftwo}%
\providecommand \href [0]{\begingroup \@sanitize@url \@href}%
\providecommand \@href[1]{\@@startlink{#1}\@@href}%
\providecommand \@@href[1]{\endgroup#1\@@endlink}%
\providecommand \@sanitize@url [0]{\catcode `\\12\catcode `\$12\catcode
  `\&12\catcode `\#12\catcode `\^12\catcode `\_12\catcode `\%12\relax}%
\providecommand \@@startlink[1]{}%
\providecommand \@@endlink[0]{}%
\providecommand \url  [0]{\begingroup\@sanitize@url \@url }%
\providecommand \@url [1]{\endgroup\@href {#1}{\urlprefix }}%
\providecommand \urlprefix  [0]{URL }%
\providecommand \Eprint [0]{\href }%
\providecommand \doibase [0]{http://dx.doi.org/}%
\providecommand \selectlanguage [0]{\@gobble}%
\providecommand \bibinfo  [0]{\@secondoftwo}%
\providecommand \bibfield  [0]{\@secondoftwo}%
\providecommand \translation [1]{[#1]}%
\providecommand \BibitemOpen [0]{}%
\providecommand \bibitemStop [0]{}%
\providecommand \bibitemNoStop [0]{.\EOS\space}%
\providecommand \EOS [0]{\spacefactor3000\relax}%
\providecommand \BibitemShut  [1]{\csname bibitem#1\endcsname}%
\let\auto@bib@innerbib\@empty
%</preamble>
\bibitem [{\citenamefont {Majorana}(1937)}]{majorana1937teoria}%
  \BibitemOpen
  \bibfield  {author} {\bibinfo {author} {\bibfnamefont {E.}~\bibnamefont
  {Majorana}},\ }\href@noop {} {\bibfield  {journal} {\bibinfo  {journal} {Il
  Nuovo Cimento (1924-1942)}\ }\textbf {\bibinfo {volume} {14}},\ \bibinfo
  {pages} {171} (\bibinfo {year} {1937})}\BibitemShut {NoStop}%
\bibitem [{\citenamefont {Alicea}(2012)}]{Alicea}%
  \BibitemOpen
  \bibfield  {author} {\bibinfo {author} {\bibfnamefont {J.}~\bibnamefont
  {Alicea}},\ }\href {\doibase 10.1088/0034-4885/75/7/076501} {\bibfield
  {journal} {\bibinfo  {journal} {Reports on Progress in Physics}\ }\textbf
  {\bibinfo {volume} {75}},\ \bibinfo {pages} {076501} (\bibinfo {year}
  {2012})}\BibitemShut {NoStop}%
\bibitem [{\citenamefont {Beenakker}(2013)}]{Beenakker2}%
  \BibitemOpen
  \bibfield  {author} {\bibinfo {author} {\bibfnamefont {C.}~\bibnamefont
  {Beenakker}},\ }\href {\doibase 10.1146/annurev-conmatphys-030212-184337}
  {\bibfield  {journal} {\bibinfo  {journal} {Annual Review of Condensed Matter
  Physics}\ }\textbf {\bibinfo {volume} {4}},\ \bibinfo {pages} {113} (\bibinfo
  {year} {2013})},\ \Eprint
  {http://arxiv.org/abs/https://doi.org/10.1146/annurev-conmatphys-030212-184337}
  {https://doi.org/10.1146/annurev-conmatphys-030212-184337} \BibitemShut
  {NoStop}%
\bibitem [{\citenamefont {Oreg}\ \emph {et~al.}(2010)\citenamefont {Oreg},
  \citenamefont {Refael},\ and\ \citenamefont {von Oppen}}]{Yoreg}%
  \BibitemOpen
  \bibfield  {author} {\bibinfo {author} {\bibfnamefont {Y.}~\bibnamefont
  {Oreg}}, \bibinfo {author} {\bibfnamefont {G.}~\bibnamefont {Refael}}, \ and\
  \bibinfo {author} {\bibfnamefont {F.}~\bibnamefont {von Oppen}},\ }\href
  {\doibase 10.1103/PhysRevLett.105.177002} {\bibfield  {journal} {\bibinfo
  {journal} {Phys. Rev. Lett.}\ }\textbf {\bibinfo {volume} {105}},\ \bibinfo
  {pages} {177002} (\bibinfo {year} {2010})}\BibitemShut {NoStop}%
\bibitem [{\citenamefont {Lutchyn}\ \emph {et~al.}(2010)\citenamefont
  {Lutchyn}, \citenamefont {Sau},\ and\ \citenamefont {Das~Sarma}}]{DasSarma}%
  \BibitemOpen
  \bibfield  {author} {\bibinfo {author} {\bibfnamefont {R.~M.}\ \bibnamefont
  {Lutchyn}}, \bibinfo {author} {\bibfnamefont {J.~D.}\ \bibnamefont {Sau}}, \
  and\ \bibinfo {author} {\bibfnamefont {S.}~\bibnamefont {Das~Sarma}},\ }\href
  {\doibase 10.1103/PhysRevLett.105.077001} {\bibfield  {journal} {\bibinfo
  {journal} {Phys. Rev. Lett.}\ }\textbf {\bibinfo {volume} {105}},\ \bibinfo
  {pages} {077001} (\bibinfo {year} {2010})}\BibitemShut {NoStop}%
\bibitem [{\citenamefont {Hasan}\ and\ \citenamefont {Kane}(2010)}]{Hasan}%
  \BibitemOpen
  \bibfield  {author} {\bibinfo {author} {\bibfnamefont {M.~Z.}\ \bibnamefont
  {Hasan}}\ and\ \bibinfo {author} {\bibfnamefont {C.~L.}\ \bibnamefont
  {Kane}},\ }\href {\doibase 10.1103/RevModPhys.82.3045} {\bibfield  {journal}
  {\bibinfo  {journal} {Rev. Mod. Phys.}\ }\textbf {\bibinfo {volume} {82}},\
  \bibinfo {pages} {3045} (\bibinfo {year} {2010})}\BibitemShut {NoStop}%
\bibitem [{\citenamefont {Sato}\ and\ \citenamefont {Fujimoto}(2016)}]{Sato}%
  \BibitemOpen
  \bibfield  {author} {\bibinfo {author} {\bibfnamefont {M.}~\bibnamefont
  {Sato}}\ and\ \bibinfo {author} {\bibfnamefont {S.}~\bibnamefont
  {Fujimoto}},\ }\href {\doibase 10.7566/JPSJ.85.072001} {\bibfield  {journal}
  {\bibinfo  {journal} {Journal of the Physical Society of Japan}\ }\textbf
  {\bibinfo {volume} {85}},\ \bibinfo {pages} {072001} (\bibinfo {year}
  {2016})},\ \Eprint
  {http://arxiv.org/abs/https://doi.org/10.7566/JPSJ.85.072001}
  {https://doi.org/10.7566/JPSJ.85.072001} \BibitemShut {NoStop}%
\bibitem [{\citenamefont {Fulga}\ \emph {et~al.}(2013)\citenamefont {Fulga},
  \citenamefont {Haim}, \citenamefont {Akhmerov},\ and\ \citenamefont
  {Oreg}}]{Fulga}%
  \BibitemOpen
  \bibfield  {author} {\bibinfo {author} {\bibfnamefont {I.~C.}\ \bibnamefont
  {Fulga}}, \bibinfo {author} {\bibfnamefont {A.}~\bibnamefont {Haim}},
  \bibinfo {author} {\bibfnamefont {A.~R.}\ \bibnamefont {Akhmerov}}, \ and\
  \bibinfo {author} {\bibfnamefont {Y.}~\bibnamefont {Oreg}},\ }\href {\doibase
  10.1088/1367-2630/15/4/045020} {\bibfield  {journal} {\bibinfo  {journal}
  {New Journal of Physics}\ }\textbf {\bibinfo {volume} {15}},\ \bibinfo
  {pages} {045020} (\bibinfo {year} {2013})}\BibitemShut {NoStop}%
\bibitem [{\citenamefont {Stanescu}\ \emph {et~al.}(2011)\citenamefont
  {Stanescu}, \citenamefont {Lutchyn},\ and\ \citenamefont
  {Das~Sarma}}]{DasSarma2011}%
  \BibitemOpen
  \bibfield  {author} {\bibinfo {author} {\bibfnamefont {T.~D.}\ \bibnamefont
  {Stanescu}}, \bibinfo {author} {\bibfnamefont {R.~M.}\ \bibnamefont
  {Lutchyn}}, \ and\ \bibinfo {author} {\bibfnamefont {S.}~\bibnamefont
  {Das~Sarma}},\ }\href {\doibase 10.1103/PhysRevB.84.144522} {\bibfield
  {journal} {\bibinfo  {journal} {Phys. Rev. B}\ }\textbf {\bibinfo {volume}
  {84}},\ \bibinfo {pages} {144522} (\bibinfo {year} {2011})}\BibitemShut
  {NoStop}%
\bibitem [{\citenamefont {Cayao}\ \emph {et~al.}(2015)\citenamefont {Cayao},
  \citenamefont {Prada}, \citenamefont {San-Jose},\ and\ \citenamefont
  {Aguado}}]{Cayao}%
  \BibitemOpen
  \bibfield  {author} {\bibinfo {author} {\bibfnamefont {J.}~\bibnamefont
  {Cayao}}, \bibinfo {author} {\bibfnamefont {E.}~\bibnamefont {Prada}},
  \bibinfo {author} {\bibfnamefont {P.}~\bibnamefont {San-Jose}}, \ and\
  \bibinfo {author} {\bibfnamefont {R.}~\bibnamefont {Aguado}},\ }\href
  {\doibase 10.1103/PhysRevB.91.024514} {\bibfield  {journal} {\bibinfo
  {journal} {Phys. Rev. B}\ }\textbf {\bibinfo {volume} {91}},\ \bibinfo
  {pages} {024514} (\bibinfo {year} {2015})}\BibitemShut {NoStop}%
\bibitem [{\citenamefont {Lu}(2020)}]{Phys13}%
  \BibitemOpen
  \bibfield  {author} {\bibinfo {author} {\bibfnamefont {H.-Z.}\ \bibnamefont
  {Lu}},\ }\href {https://physics.aps.org/articles/v13/30} {\bibfield
  {journal} {\bibinfo  {journal} {Physics}\ }\textbf {\bibinfo {volume} {13}},\
  \bibinfo {pages} {30} (\bibinfo {year} {2020})}\BibitemShut {NoStop}%
\bibitem [{\citenamefont {Mourik}\ \emph {et~al.}(2012)\citenamefont {Mourik},
  \citenamefont {Zuo}, \citenamefont {Frolov}, \citenamefont {Plissard},
  \citenamefont {Bakkers},\ and\ \citenamefont {Kouwenhoven}}]{Mourik}%
  \BibitemOpen
  \bibfield  {author} {\bibinfo {author} {\bibfnamefont {V.}~\bibnamefont
  {Mourik}}, \bibinfo {author} {\bibfnamefont {K.}~\bibnamefont {Zuo}},
  \bibinfo {author} {\bibfnamefont {S.~M.}\ \bibnamefont {Frolov}}, \bibinfo
  {author} {\bibfnamefont {S.~R.}\ \bibnamefont {Plissard}}, \bibinfo {author}
  {\bibfnamefont {E.~P. A.~M.}\ \bibnamefont {Bakkers}}, \ and\ \bibinfo
  {author} {\bibfnamefont {L.~P.}\ \bibnamefont {Kouwenhoven}},\ }\href
  {\doibase 10.1126/science.1222360} {\bibfield  {journal} {\bibinfo  {journal}
  {Science}\ }\textbf {\bibinfo {volume} {336}},\ \bibinfo {pages} {1003}
  (\bibinfo {year} {2012})},\ \Eprint
  {http://arxiv.org/abs/http://science.sciencemag.org/content/336/6084/1003.full.pdf}
  {http://science.sciencemag.org/content/336/6084/1003.full.pdf} \BibitemShut
  {NoStop}%
\bibitem [{\citenamefont {Das}\ \emph {et~al.}(2012)\citenamefont {Das},
  \citenamefont {Ronen}, \citenamefont {Most}, \citenamefont {Oreg},
  \citenamefont {Heiblum},\ and\ \citenamefont {Shtrikman}}]{Das}%
  \BibitemOpen
  \bibfield  {author} {\bibinfo {author} {\bibfnamefont {A.}~\bibnamefont
  {Das}}, \bibinfo {author} {\bibfnamefont {Y.}~\bibnamefont {Ronen}}, \bibinfo
  {author} {\bibfnamefont {Y.}~\bibnamefont {Most}}, \bibinfo {author}
  {\bibfnamefont {Y.}~\bibnamefont {Oreg}}, \bibinfo {author} {\bibfnamefont
  {M.}~\bibnamefont {Heiblum}}, \ and\ \bibinfo {author} {\bibfnamefont
  {H.}~\bibnamefont {Shtrikman}},\ }\href {https://doi.org/10.1038/nphys2479}
  {\bibfield  {journal} {\bibinfo  {journal} {Nature Physics}\ }\textbf
  {\bibinfo {volume} {8}},\ \bibinfo {pages} {887 EP } (\bibinfo {year}
  {2012})}\BibitemShut {NoStop}%
\bibitem [{\citenamefont {Finck}\ \emph {et~al.}(2013)\citenamefont {Finck},
  \citenamefont {Van~Harlingen}, \citenamefont {Mohseni}, \citenamefont
  {Jung},\ and\ \citenamefont {Li}}]{Finck}%
  \BibitemOpen
  \bibfield  {author} {\bibinfo {author} {\bibfnamefont {A.~D.~K.}\
  \bibnamefont {Finck}}, \bibinfo {author} {\bibfnamefont {D.~J.}\ \bibnamefont
  {Van~Harlingen}}, \bibinfo {author} {\bibfnamefont {P.~K.}\ \bibnamefont
  {Mohseni}}, \bibinfo {author} {\bibfnamefont {K.}~\bibnamefont {Jung}}, \
  and\ \bibinfo {author} {\bibfnamefont {X.}~\bibnamefont {Li}},\ }\href
  {\doibase 10.1103/PhysRevLett.110.126406} {\bibfield  {journal} {\bibinfo
  {journal} {Phys. Rev. Lett.}\ }\textbf {\bibinfo {volume} {110}},\ \bibinfo
  {pages} {126406} (\bibinfo {year} {2013})}\BibitemShut {NoStop}%
\bibitem [{\citenamefont {Rokhinson}\ \emph {et~al.}(2012)\citenamefont
  {Rokhinson}, \citenamefont {Liu},\ and\ \citenamefont {Furdyna}}]{Rokhinson}%
  \BibitemOpen
  \bibfield  {author} {\bibinfo {author} {\bibfnamefont {L.~P.}\ \bibnamefont
  {Rokhinson}}, \bibinfo {author} {\bibfnamefont {X.}~\bibnamefont {Liu}}, \
  and\ \bibinfo {author} {\bibfnamefont {J.~K.}\ \bibnamefont {Furdyna}},\
  }\href {https://doi.org/10.1038/nphys2429} {\bibfield  {journal} {\bibinfo
  {journal} {Nature Physics}\ }\textbf {\bibinfo {volume} {8}},\ \bibinfo
  {pages} {795 EP } (\bibinfo {year} {2012})}\BibitemShut {NoStop}%
\bibitem [{\citenamefont {Nichele}\ \emph {et~al.}(2017)\citenamefont
  {Nichele}, \citenamefont {Drachmann}, \citenamefont {Whiticar}, \citenamefont
  {O'Farrell}, \citenamefont {Suominen}, \citenamefont {Fornieri},
  \citenamefont {Wang}, \citenamefont {Gardner}, \citenamefont {Thomas},
  \citenamefont {Hatke}, \citenamefont {Krogstrup}, \citenamefont {Manfra},
  \citenamefont {Flensberg},\ and\ \citenamefont {Marcus}}]{Nichele}%
  \BibitemOpen
  \bibfield  {author} {\bibinfo {author} {\bibfnamefont {F.}~\bibnamefont
  {Nichele}}, \bibinfo {author} {\bibfnamefont {A.~C.~C.}\ \bibnamefont
  {Drachmann}}, \bibinfo {author} {\bibfnamefont {A.~M.}\ \bibnamefont
  {Whiticar}}, \bibinfo {author} {\bibfnamefont {E.~C.~T.}\ \bibnamefont
  {O'Farrell}}, \bibinfo {author} {\bibfnamefont {H.~J.}\ \bibnamefont
  {Suominen}}, \bibinfo {author} {\bibfnamefont {A.}~\bibnamefont {Fornieri}},
  \bibinfo {author} {\bibfnamefont {T.}~\bibnamefont {Wang}}, \bibinfo {author}
  {\bibfnamefont {G.~C.}\ \bibnamefont {Gardner}}, \bibinfo {author}
  {\bibfnamefont {C.}~\bibnamefont {Thomas}}, \bibinfo {author} {\bibfnamefont
  {A.~T.}\ \bibnamefont {Hatke}}, \bibinfo {author} {\bibfnamefont
  {P.}~\bibnamefont {Krogstrup}}, \bibinfo {author} {\bibfnamefont {M.~J.}\
  \bibnamefont {Manfra}}, \bibinfo {author} {\bibfnamefont {K.}~\bibnamefont
  {Flensberg}}, \ and\ \bibinfo {author} {\bibfnamefont {C.~M.}\ \bibnamefont
  {Marcus}},\ }\href {\doibase 10.1103/PhysRevLett.119.136803} {\bibfield
  {journal} {\bibinfo  {journal} {Phys. Rev. Lett.}\ }\textbf {\bibinfo
  {volume} {119}},\ \bibinfo {pages} {136803} (\bibinfo {year}
  {2017})}\BibitemShut {NoStop}%
\bibitem [{\citenamefont {Flensberg}(2010)}]{Flensberg}%
  \BibitemOpen
  \bibfield  {author} {\bibinfo {author} {\bibfnamefont {K.}~\bibnamefont
  {Flensberg}},\ }\href {\doibase 10.1103/PhysRevB.82.180516} {\bibfield
  {journal} {\bibinfo  {journal} {Phys. Rev. B}\ }\textbf {\bibinfo {volume}
  {82}},\ \bibinfo {pages} {180516} (\bibinfo {year} {2010})}\BibitemShut
  {NoStop}%
\bibitem [{\citenamefont {Liu}\ \emph {et~al.}(2012)\citenamefont {Liu},
  \citenamefont {Potter}, \citenamefont {Law},\ and\ \citenamefont
  {Lee}}]{Multiband}%
  \BibitemOpen
  \bibfield  {author} {\bibinfo {author} {\bibfnamefont {J.}~\bibnamefont
  {Liu}}, \bibinfo {author} {\bibfnamefont {A.~C.}\ \bibnamefont {Potter}},
  \bibinfo {author} {\bibfnamefont {K.~T.}\ \bibnamefont {Law}}, \ and\
  \bibinfo {author} {\bibfnamefont {P.~A.}\ \bibnamefont {Lee}},\ }\href
  {\doibase 10.1103/PhysRevLett.109.267002} {\bibfield  {journal} {\bibinfo
  {journal} {Phys. Rev. Lett.}\ }\textbf {\bibinfo {volume} {109}},\ \bibinfo
  {pages} {267002} (\bibinfo {year} {2012})}\BibitemShut {NoStop}%
\bibitem [{\citenamefont {Pikulin}\ \emph {et~al.}(2012)\citenamefont
  {Pikulin}, \citenamefont {Dahlhaus}, \citenamefont {Wimmer}, \citenamefont
  {Schomerus},\ and\ \citenamefont {Beenakker}}]{weak_anti}%
  \BibitemOpen
  \bibfield  {author} {\bibinfo {author} {\bibfnamefont {D.~I.}\ \bibnamefont
  {Pikulin}}, \bibinfo {author} {\bibfnamefont {J.~P.}\ \bibnamefont
  {Dahlhaus}}, \bibinfo {author} {\bibfnamefont {M.}~\bibnamefont {Wimmer}},
  \bibinfo {author} {\bibfnamefont {H.}~\bibnamefont {Schomerus}}, \ and\
  \bibinfo {author} {\bibfnamefont {C.~W.~J.}\ \bibnamefont {Beenakker}},\
  }\href {\doibase 10.1088/1367-2630/14/12/125011} {\bibfield  {journal}
  {\bibinfo  {journal} {New Journal of Physics}\ }\textbf {\bibinfo {volume}
  {14}},\ \bibinfo {pages} {125011} (\bibinfo {year} {2012})}\BibitemShut
  {NoStop}%
\bibitem [{\citenamefont {Goldhaber-Gordon}\ \emph {et~al.}(1998)\citenamefont
  {Goldhaber-Gordon}, \citenamefont {Shtrikman}, \citenamefont {Mahalu},
  \citenamefont {Abusch-Magder}, \citenamefont {Meirav},\ and\ \citenamefont
  {Kastner}}]{goldhaber1998kondo}%
  \BibitemOpen
  \bibfield  {author} {\bibinfo {author} {\bibfnamefont {D.}~\bibnamefont
  {Goldhaber-Gordon}}, \bibinfo {author} {\bibfnamefont {H.}~\bibnamefont
  {Shtrikman}}, \bibinfo {author} {\bibfnamefont {D.}~\bibnamefont {Mahalu}},
  \bibinfo {author} {\bibfnamefont {D.}~\bibnamefont {Abusch-Magder}}, \bibinfo
  {author} {\bibfnamefont {U.}~\bibnamefont {Meirav}}, \ and\ \bibinfo {author}
  {\bibfnamefont {M.}~\bibnamefont {Kastner}},\ }\href@noop {} {\bibfield
  {journal} {\bibinfo  {journal} {Nature}\ }\textbf {\bibinfo {volume} {391}},\
  \bibinfo {pages} {156} (\bibinfo {year} {1998})}\BibitemShut {NoStop}%
\bibitem [{\citenamefont {Hell}\ \emph {et~al.}(2018)\citenamefont {Hell},
  \citenamefont {Flensberg},\ and\ \citenamefont {Leijnse}}]{Hell}%
  \BibitemOpen
  \bibfield  {author} {\bibinfo {author} {\bibfnamefont {M.}~\bibnamefont
  {Hell}}, \bibinfo {author} {\bibfnamefont {K.}~\bibnamefont {Flensberg}}, \
  and\ \bibinfo {author} {\bibfnamefont {M.}~\bibnamefont {Leijnse}},\ }\href
  {\doibase 10.1103/PhysRevB.97.161401} {\bibfield  {journal} {\bibinfo
  {journal} {Phys. Rev. B}\ }\textbf {\bibinfo {volume} {97}},\ \bibinfo
  {pages} {161401} (\bibinfo {year} {2018})}\BibitemShut {NoStop}%
\bibitem [{\citenamefont {Tripathi}\ \emph {et~al.}(2016)\citenamefont
  {Tripathi}, \citenamefont {Das},\ and\ \citenamefont {Rao}}]{Fingerprints}%
  \BibitemOpen
  \bibfield  {author} {\bibinfo {author} {\bibfnamefont {K.~M.}\ \bibnamefont
  {Tripathi}}, \bibinfo {author} {\bibfnamefont {S.}~\bibnamefont {Das}}, \
  and\ \bibinfo {author} {\bibfnamefont {S.}~\bibnamefont {Rao}},\ }\href
  {\doibase 10.1103/PhysRevLett.116.166401} {\bibfield  {journal} {\bibinfo
  {journal} {Phys. Rev. Lett.}\ }\textbf {\bibinfo {volume} {116}},\ \bibinfo
  {pages} {166401} (\bibinfo {year} {2016})}\BibitemShut {NoStop}%
\bibitem [{\citenamefont {Haim}\ \emph {et~al.}(2015)\citenamefont {Haim},
  \citenamefont {Berg}, \citenamefont {von Oppen},\ and\ \citenamefont
  {Oreg}}]{Yoreg1}%
  \BibitemOpen
  \bibfield  {author} {\bibinfo {author} {\bibfnamefont {A.}~\bibnamefont
  {Haim}}, \bibinfo {author} {\bibfnamefont {E.}~\bibnamefont {Berg}}, \bibinfo
  {author} {\bibfnamefont {F.}~\bibnamefont {von Oppen}}, \ and\ \bibinfo
  {author} {\bibfnamefont {Y.}~\bibnamefont {Oreg}},\ }\href {\doibase
  10.1103/PhysRevLett.114.166406} {\bibfield  {journal} {\bibinfo  {journal}
  {Phys. Rev. Lett.}\ }\textbf {\bibinfo {volume} {114}},\ \bibinfo {pages}
  {166406} (\bibinfo {year} {2015})}\BibitemShut {NoStop}%
\bibitem [{\citenamefont {Liu}\ \emph {et~al.}(2018)\citenamefont {Liu},
  \citenamefont {Sau},\ and\ \citenamefont {Das~Sarma}}]{DasSarma1}%
  \BibitemOpen
  \bibfield  {author} {\bibinfo {author} {\bibfnamefont {C.-X.}\ \bibnamefont
  {Liu}}, \bibinfo {author} {\bibfnamefont {J.~D.}\ \bibnamefont {Sau}}, \ and\
  \bibinfo {author} {\bibfnamefont {S.}~\bibnamefont {Das~Sarma}},\ }\href
  {\doibase 10.1103/PhysRevB.97.214502} {\bibfield  {journal} {\bibinfo
  {journal} {Phys. Rev. B}\ }\textbf {\bibinfo {volume} {97}},\ \bibinfo
  {pages} {214502} (\bibinfo {year} {2018})}\BibitemShut {NoStop}%
\bibitem [{\citenamefont {Liu}\ \emph {et~al.}(2017)\citenamefont {Liu},
  \citenamefont {Sau}, \citenamefont {Stanescu},\ and\ \citenamefont
  {Das~Sarma}}]{DasSarma2}%
  \BibitemOpen
  \bibfield  {author} {\bibinfo {author} {\bibfnamefont {C.-X.}\ \bibnamefont
  {Liu}}, \bibinfo {author} {\bibfnamefont {J.~D.}\ \bibnamefont {Sau}},
  \bibinfo {author} {\bibfnamefont {T.~D.}\ \bibnamefont {Stanescu}}, \ and\
  \bibinfo {author} {\bibfnamefont {S.}~\bibnamefont {Das~Sarma}},\ }\href
  {\doibase 10.1103/PhysRevB.96.075161} {\bibfield  {journal} {\bibinfo
  {journal} {Phys. Rev. B}\ }\textbf {\bibinfo {volume} {96}},\ \bibinfo
  {pages} {075161} (\bibinfo {year} {2017})}\BibitemShut {NoStop}%
\bibitem [{\citenamefont {Ricco}\ \emph
  {et~al.}(2018{\natexlab{a}})\citenamefont {Ricco}, \citenamefont {de~Souza},
  \citenamefont {Figueira}, \citenamefont {Shelykh},\ and\ \citenamefont
  {Seridonio}}]{Seridonio}%
  \BibitemOpen
  \bibfield  {author} {\bibinfo {author} {\bibfnamefont {L.}~\bibnamefont
  {Ricco}}, \bibinfo {author} {\bibfnamefont {M.}~\bibnamefont {de~Souza}},
  \bibinfo {author} {\bibfnamefont {M.}~\bibnamefont {Figueira}}, \bibinfo
  {author} {\bibfnamefont {I.}~\bibnamefont {Shelykh}}, \ and\ \bibinfo
  {author} {\bibfnamefont {A.}~\bibnamefont {Seridonio}},\ }\href@noop {}
  {\bibfield  {journal} {\bibinfo  {journal} {arXiv preprint arXiv:1811.10305}\
  } (\bibinfo {year} {2018}{\natexlab{a}})}\BibitemShut {NoStop}%
\bibitem [{\citenamefont {Deng}\ \emph {et~al.}(2018)\citenamefont {Deng},
  \citenamefont {Vaitiek\ifmmode~\dot{e}\else \.{e}\fi{}nas}, \citenamefont
  {Prada}, \citenamefont {San-Jose}, \citenamefont {Nyg\aa{}rd}, \citenamefont
  {Krogstrup}, \citenamefont {Aguado},\ and\ \citenamefont
  {Marcus}}]{Deng2018}%
  \BibitemOpen
  \bibfield  {author} {\bibinfo {author} {\bibfnamefont {M.-T.}\ \bibnamefont
  {Deng}}, \bibinfo {author} {\bibfnamefont {S.}~\bibnamefont
  {Vaitiek\ifmmode~\dot{e}\else \.{e}\fi{}nas}}, \bibinfo {author}
  {\bibfnamefont {E.}~\bibnamefont {Prada}}, \bibinfo {author} {\bibfnamefont
  {P.}~\bibnamefont {San-Jose}}, \bibinfo {author} {\bibfnamefont
  {J.}~\bibnamefont {Nyg\aa{}rd}}, \bibinfo {author} {\bibfnamefont
  {P.}~\bibnamefont {Krogstrup}}, \bibinfo {author} {\bibfnamefont
  {R.}~\bibnamefont {Aguado}}, \ and\ \bibinfo {author} {\bibfnamefont {C.~M.}\
  \bibnamefont {Marcus}},\ }\href {\doibase 10.1103/PhysRevB.98.085125}
  {\bibfield  {journal} {\bibinfo  {journal} {Phys. Rev. B}\ }\textbf {\bibinfo
  {volume} {98}},\ \bibinfo {pages} {085125} (\bibinfo {year}
  {2018})}\BibitemShut {NoStop}%
\bibitem [{\citenamefont {Sau}\ \emph {et~al.}(2015)\citenamefont {Sau},
  \citenamefont {Swingle},\ and\ \citenamefont {Tewari}}]{Sau}%
  \BibitemOpen
  \bibfield  {author} {\bibinfo {author} {\bibfnamefont {J.~D.}\ \bibnamefont
  {Sau}}, \bibinfo {author} {\bibfnamefont {B.}~\bibnamefont {Swingle}}, \ and\
  \bibinfo {author} {\bibfnamefont {S.}~\bibnamefont {Tewari}},\ }\href
  {\doibase 10.1103/PhysRevB.92.020511} {\bibfield  {journal} {\bibinfo
  {journal} {Phys. Rev. B}\ }\textbf {\bibinfo {volume} {92}},\ \bibinfo
  {pages} {020511} (\bibinfo {year} {2015})}\BibitemShut {NoStop}%
\bibitem [{\citenamefont {Liu}\ and\ \citenamefont
  {Baranger}(2011)}]{PhysRevB.84.201308}%
  \BibitemOpen
  \bibfield  {author} {\bibinfo {author} {\bibfnamefont {D.~E.}\ \bibnamefont
  {Liu}}\ and\ \bibinfo {author} {\bibfnamefont {H.~U.}\ \bibnamefont
  {Baranger}},\ }\href {\doibase 10.1103/PhysRevB.84.201308} {\bibfield
  {journal} {\bibinfo  {journal} {Phys. Rev. B}\ }\textbf {\bibinfo {volume}
  {84}},\ \bibinfo {pages} {201308} (\bibinfo {year} {2011})}\BibitemShut
  {NoStop}%
\bibitem [{\citenamefont {Cao}\ \emph {et~al.}(2012)\citenamefont {Cao},
  \citenamefont {Wang}, \citenamefont {Xiong}, \citenamefont {Gong},\ and\
  \citenamefont {Li}}]{Cao2012}%
  \BibitemOpen
  \bibfield  {author} {\bibinfo {author} {\bibfnamefont {Y.}~\bibnamefont
  {Cao}}, \bibinfo {author} {\bibfnamefont {P.}~\bibnamefont {Wang}}, \bibinfo
  {author} {\bibfnamefont {G.}~\bibnamefont {Xiong}}, \bibinfo {author}
  {\bibfnamefont {M.}~\bibnamefont {Gong}}, \ and\ \bibinfo {author}
  {\bibfnamefont {X.-Q.}\ \bibnamefont {Li}},\ }\href {\doibase
  10.1103/PhysRevB.86.115311} {\bibfield  {journal} {\bibinfo  {journal} {Phys.
  Rev. B}\ }\textbf {\bibinfo {volume} {86}},\ \bibinfo {pages} {115311}
  (\bibinfo {year} {2012})}\BibitemShut {NoStop}%
\bibitem [{\citenamefont {Ricco}\ \emph
  {et~al.}(2018{\natexlab{b}})\citenamefont {Ricco}, \citenamefont {Dessotti},
  \citenamefont {Shelykh}, \citenamefont {Figueira},\ and\ \citenamefont
  {Seridonio}}]{ricco2018tuning}%
  \BibitemOpen
  \bibfield  {author} {\bibinfo {author} {\bibfnamefont {L.}~\bibnamefont
  {Ricco}}, \bibinfo {author} {\bibfnamefont {F.}~\bibnamefont {Dessotti}},
  \bibinfo {author} {\bibfnamefont {I.}~\bibnamefont {Shelykh}}, \bibinfo
  {author} {\bibfnamefont {M.}~\bibnamefont {Figueira}}, \ and\ \bibinfo
  {author} {\bibfnamefont {A.}~\bibnamefont {Seridonio}},\ }\href@noop {}
  {\bibfield  {journal} {\bibinfo  {journal} {Scientific reports}\ }\textbf
  {\bibinfo {volume} {8}},\ \bibinfo {pages} {2790} (\bibinfo {year}
  {2018}{\natexlab{b}})}\BibitemShut {NoStop}%
\bibitem [{\citenamefont {Leijnse}(2014)}]{Leijnse_2014}%
  \BibitemOpen
  \bibfield  {author} {\bibinfo {author} {\bibfnamefont {M.}~\bibnamefont
  {Leijnse}},\ }\href {\doibase 10.1088/1367-2630/16/1/015029} {\bibfield
  {journal} {\bibinfo  {journal} {New Journal of Physics}\ }\textbf {\bibinfo
  {volume} {16}},\ \bibinfo {pages} {015029} (\bibinfo {year}
  {2014})}\BibitemShut {NoStop}%
\bibitem [{\citenamefont {Dom\'{\i}nguez}\ \emph {et~al.}(2012)\citenamefont
  {Dom\'{\i}nguez}, \citenamefont {Hassler},\ and\ \citenamefont
  {Platero}}]{PhysRevB.86.140503}%
  \BibitemOpen
  \bibfield  {author} {\bibinfo {author} {\bibfnamefont {F.}~\bibnamefont
  {Dom\'{\i}nguez}}, \bibinfo {author} {\bibfnamefont {F.}~\bibnamefont
  {Hassler}}, \ and\ \bibinfo {author} {\bibfnamefont {G.}~\bibnamefont
  {Platero}},\ }\href {\doibase 10.1103/PhysRevB.86.140503} {\bibfield
  {journal} {\bibinfo  {journal} {Phys. Rev. B}\ }\textbf {\bibinfo {volume}
  {86}},\ \bibinfo {pages} {140503} (\bibinfo {year} {2012})}\BibitemShut
  {NoStop}%
\bibitem [{\citenamefont {San-Jose}\ \emph {et~al.}(2012)\citenamefont
  {San-Jose}, \citenamefont {Prada},\ and\ \citenamefont
  {Aguado}}]{PhysRevLett.108.257001}%
  \BibitemOpen
  \bibfield  {author} {\bibinfo {author} {\bibfnamefont {P.}~\bibnamefont
  {San-Jose}}, \bibinfo {author} {\bibfnamefont {E.}~\bibnamefont {Prada}}, \
  and\ \bibinfo {author} {\bibfnamefont {R.}~\bibnamefont {Aguado}},\ }\href
  {\doibase 10.1103/PhysRevLett.108.257001} {\bibfield  {journal} {\bibinfo
  {journal} {Phys. Rev. Lett.}\ }\textbf {\bibinfo {volume} {108}},\ \bibinfo
  {pages} {257001} (\bibinfo {year} {2012})}\BibitemShut {NoStop}%
\bibitem [{\citenamefont {L\"u}\ \emph {et~al.}(2016)\citenamefont {L\"u},
  \citenamefont {Lu},\ and\ \citenamefont {Shen}}]{Lu2016}%
  \BibitemOpen
  \bibfield  {author} {\bibinfo {author} {\bibfnamefont {H.-F.}\ \bibnamefont
  {L\"u}}, \bibinfo {author} {\bibfnamefont {H.-Z.}\ \bibnamefont {Lu}}, \ and\
  \bibinfo {author} {\bibfnamefont {S.-Q.}\ \bibnamefont {Shen}},\ }\href
  {\doibase 10.1103/PhysRevB.93.245418} {\bibfield  {journal} {\bibinfo
  {journal} {Phys. Rev. B}\ }\textbf {\bibinfo {volume} {93}},\ \bibinfo
  {pages} {245418} (\bibinfo {year} {2016})}\BibitemShut {NoStop}%
\bibitem [{\citenamefont {Chen}\ \emph {et~al.}(2014)\citenamefont {Chen},
  \citenamefont {Chen},\ and\ \citenamefont {Zhao}}]{Chen_2014}%
  \BibitemOpen
  \bibfield  {author} {\bibinfo {author} {\bibfnamefont {Q.}~\bibnamefont
  {Chen}}, \bibinfo {author} {\bibfnamefont {K.-Q.}\ \bibnamefont {Chen}}, \
  and\ \bibinfo {author} {\bibfnamefont {H.-K.}\ \bibnamefont {Zhao}},\ }\href
  {\doibase 10.1088/0953-8984/26/31/315011} {\bibfield  {journal} {\bibinfo
  {journal} {Journal of Physics: Condensed Matter}\ }\textbf {\bibinfo {volume}
  {26}},\ \bibinfo {pages} {315011} (\bibinfo {year} {2014})}\BibitemShut
  {NoStop}%
\bibitem [{\citenamefont {Devillard}\ \emph {et~al.}(2017)\citenamefont
  {Devillard}, \citenamefont {Chevallier},\ and\ \citenamefont
  {Albert}}]{PhysRevB.96.115413}%
  \BibitemOpen
  \bibfield  {author} {\bibinfo {author} {\bibfnamefont {P.}~\bibnamefont
  {Devillard}}, \bibinfo {author} {\bibfnamefont {D.}~\bibnamefont
  {Chevallier}}, \ and\ \bibinfo {author} {\bibfnamefont {M.}~\bibnamefont
  {Albert}},\ }\href {\doibase 10.1103/PhysRevB.96.115413} {\bibfield
  {journal} {\bibinfo  {journal} {Phys. Rev. B}\ }\textbf {\bibinfo {volume}
  {96}},\ \bibinfo {pages} {115413} (\bibinfo {year} {2017})}\BibitemShut
  {NoStop}%
\bibitem [{\citenamefont {Jonckheere}\ \emph {et~al.}(2019)\citenamefont
  {Jonckheere}, \citenamefont {Rech}, \citenamefont {Zazunov}, \citenamefont
  {Egger}, \citenamefont {Yeyati},\ and\ \citenamefont
  {Martin}}]{PhysRevLett.122.097003}%
  \BibitemOpen
  \bibfield  {author} {\bibinfo {author} {\bibfnamefont {T.}~\bibnamefont
  {Jonckheere}}, \bibinfo {author} {\bibfnamefont {J.}~\bibnamefont {Rech}},
  \bibinfo {author} {\bibfnamefont {A.}~\bibnamefont {Zazunov}}, \bibinfo
  {author} {\bibfnamefont {R.}~\bibnamefont {Egger}}, \bibinfo {author}
  {\bibfnamefont {A.~L.}\ \bibnamefont {Yeyati}}, \ and\ \bibinfo {author}
  {\bibfnamefont {T.}~\bibnamefont {Martin}},\ }\href {\doibase
  10.1103/PhysRevLett.122.097003} {\bibfield  {journal} {\bibinfo  {journal}
  {Phys. Rev. Lett.}\ }\textbf {\bibinfo {volume} {122}},\ \bibinfo {pages}
  {097003} (\bibinfo {year} {2019})}\BibitemShut {NoStop}%
\bibitem [{\citenamefont {Liu}\ \emph {et~al.}(2015{\natexlab{a}})\citenamefont
  {Liu}, \citenamefont {Cheng},\ and\ \citenamefont
  {Lutchyn}}]{PhysRevB.91.081405}%
  \BibitemOpen
  \bibfield  {author} {\bibinfo {author} {\bibfnamefont {D.~E.}\ \bibnamefont
  {Liu}}, \bibinfo {author} {\bibfnamefont {M.}~\bibnamefont {Cheng}}, \ and\
  \bibinfo {author} {\bibfnamefont {R.~M.}\ \bibnamefont {Lutchyn}},\ }\href
  {\doibase 10.1103/PhysRevB.91.081405} {\bibfield  {journal} {\bibinfo
  {journal} {Phys. Rev. B}\ }\textbf {\bibinfo {volume} {91}},\ \bibinfo
  {pages} {081405} (\bibinfo {year} {2015}{\natexlab{a}})}\BibitemShut
  {NoStop}%
\bibitem [{\citenamefont {Schuray}\ \emph {et~al.}(2020)\citenamefont
  {Schuray}, \citenamefont {Rammler},\ and\ \citenamefont
  {Recher}}]{PhysRevB.102.045303}%
  \BibitemOpen
  \bibfield  {author} {\bibinfo {author} {\bibfnamefont {A.}~\bibnamefont
  {Schuray}}, \bibinfo {author} {\bibfnamefont {M.}~\bibnamefont {Rammler}}, \
  and\ \bibinfo {author} {\bibfnamefont {P.}~\bibnamefont {Recher}},\ }\href
  {\doibase 10.1103/PhysRevB.102.045303} {\bibfield  {journal} {\bibinfo
  {journal} {Phys. Rev. B}\ }\textbf {\bibinfo {volume} {102}},\ \bibinfo
  {pages} {045303} (\bibinfo {year} {2020})}\BibitemShut {NoStop}%
\bibitem [{\citenamefont {Manousakis}\ \emph {et~al.}(2020)\citenamefont
  {Manousakis}, \citenamefont {Wille}, \citenamefont {Altland}, \citenamefont
  {Egger}, \citenamefont {Flensberg},\ and\ \citenamefont
  {Hassler}}]{PhysRevLett.124.096801}%
  \BibitemOpen
  \bibfield  {author} {\bibinfo {author} {\bibfnamefont {J.}~\bibnamefont
  {Manousakis}}, \bibinfo {author} {\bibfnamefont {C.}~\bibnamefont {Wille}},
  \bibinfo {author} {\bibfnamefont {A.}~\bibnamefont {Altland}}, \bibinfo
  {author} {\bibfnamefont {R.}~\bibnamefont {Egger}}, \bibinfo {author}
  {\bibfnamefont {K.}~\bibnamefont {Flensberg}}, \ and\ \bibinfo {author}
  {\bibfnamefont {F.}~\bibnamefont {Hassler}},\ }\href {\doibase
  10.1103/PhysRevLett.124.096801} {\bibfield  {journal} {\bibinfo  {journal}
  {Phys. Rev. Lett.}\ }\textbf {\bibinfo {volume} {124}},\ \bibinfo {pages}
  {096801} (\bibinfo {year} {2020})}\BibitemShut {NoStop}%
\bibitem [{\citenamefont {Guerci}\ and\ \citenamefont
  {Nava}(2019)}]{guerci2019probing}%
  \BibitemOpen
  \bibfield  {author} {\bibinfo {author} {\bibfnamefont {D.}~\bibnamefont
  {Guerci}}\ and\ \bibinfo {author} {\bibfnamefont {A.}~\bibnamefont {Nava}},\
  }\href@noop {} {\bibfield  {journal} {\bibinfo  {journal} {arXiv preprint
  arXiv:1907.06444}\ } (\bibinfo {year} {2019})}\BibitemShut {NoStop}%
\bibitem [{\citenamefont {Cheng}\ \emph {et~al.}(2014)\citenamefont {Cheng},
  \citenamefont {Becker}, \citenamefont {Bauer},\ and\ \citenamefont
  {Lutchyn}}]{PhysRevX.4.031051}%
  \BibitemOpen
  \bibfield  {author} {\bibinfo {author} {\bibfnamefont {M.}~\bibnamefont
  {Cheng}}, \bibinfo {author} {\bibfnamefont {M.}~\bibnamefont {Becker}},
  \bibinfo {author} {\bibfnamefont {B.}~\bibnamefont {Bauer}}, \ and\ \bibinfo
  {author} {\bibfnamefont {R.~M.}\ \bibnamefont {Lutchyn}},\ }\href {\doibase
  10.1103/PhysRevX.4.031051} {\bibfield  {journal} {\bibinfo  {journal} {Phys.
  Rev. X}\ }\textbf {\bibinfo {volume} {4}},\ \bibinfo {pages} {031051}
  (\bibinfo {year} {2014})}\BibitemShut {NoStop}%
\bibitem [{\citenamefont {Chiu}\ \emph {et~al.}(2018)\citenamefont {Chiu},
  \citenamefont {Sau},\ and\ \citenamefont {Das~Sarma}}]{PhysRevB.97.035310}%
  \BibitemOpen
  \bibfield  {author} {\bibinfo {author} {\bibfnamefont {C.-K.}\ \bibnamefont
  {Chiu}}, \bibinfo {author} {\bibfnamefont {J.~D.}\ \bibnamefont {Sau}}, \
  and\ \bibinfo {author} {\bibfnamefont {S.}~\bibnamefont {Das~Sarma}},\ }\href
  {\doibase 10.1103/PhysRevB.97.035310} {\bibfield  {journal} {\bibinfo
  {journal} {Phys. Rev. B}\ }\textbf {\bibinfo {volume} {97}},\ \bibinfo
  {pages} {035310} (\bibinfo {year} {2018})}\BibitemShut {NoStop}%
\bibitem [{\citenamefont {Haug}\ and\ \citenamefont
  {Jauho}(2008)}]{Jauho_book}%
  \BibitemOpen
  \bibfield  {author} {\bibinfo {author} {\bibfnamefont {H.}~\bibnamefont
  {Haug}}\ and\ \bibinfo {author} {\bibfnamefont {A.-P.}\ \bibnamefont
  {Jauho}},\ }\href@noop {} {\emph {\bibinfo {title} {Quantum kinetics in
  transport and optics of semiconductors}}},\ Vol.~\bibinfo {volume} {2}\
  (\bibinfo  {publisher} {Springer},\ \bibinfo {year} {2008})\BibitemShut
  {NoStop}%
\bibitem [{\citenamefont {Blanter}\ and\ \citenamefont
  {B{\"u}ttiker}(2000)}]{blanter2000shot_footnote}%
  \BibitemOpen
  \bibfield  {author} {\bibinfo {author} {\bibfnamefont {Y.~M.}\ \bibnamefont
  {Blanter}}\ and\ \bibinfo {author} {\bibfnamefont {M.}~\bibnamefont
  {B{\"u}ttiker}},\ }\href@noop {} {\bibfield  {journal} {\bibinfo  {journal}
  {Physics reports}\ }\textbf {\bibinfo {volume} {336}},\ \bibinfo {pages} {1}
  (\bibinfo {year} {2000})}\BibitemShut {NoStop}%
\bibitem [{\citenamefont {Zeng}\ \emph {et~al.}(2016)\citenamefont {Zeng},
  \citenamefont {Chen}, \citenamefont {You},\ and\ \citenamefont
  {L{\"u}}}]{Zeng2017}%
  \BibitemOpen
  \bibfield  {author} {\bibinfo {author} {\bibfnamefont {Q.-B.}\ \bibnamefont
  {Zeng}}, \bibinfo {author} {\bibfnamefont {S.}~\bibnamefont {Chen}}, \bibinfo
  {author} {\bibfnamefont {L.}~\bibnamefont {You}}, \ and\ \bibinfo {author}
  {\bibfnamefont {R.}~\bibnamefont {L{\"u}}},\ }\href {\doibase
  10.1007/s11467-016-0620-3} {\bibfield  {journal} {\bibinfo  {journal}
  {Frontiers of Physics}\ ,\ \bibinfo {pages} {127302}} (\bibinfo {year}
  {2016})}\BibitemShut {NoStop}%
\bibitem [{\citenamefont {Calle}\ \emph {et~al.}(2020)\citenamefont {Calle},
  \citenamefont {Pacheco}, \citenamefont {Orellana},\ and\ \citenamefont
  {Ot{\'a}lora}}]{AMCalle_Correspondence}%
  \BibitemOpen
  \bibfield  {author} {\bibinfo {author} {\bibfnamefont {A.~M.}\ \bibnamefont
  {Calle}}, \bibinfo {author} {\bibfnamefont {M.}~\bibnamefont {Pacheco}},
  \bibinfo {author} {\bibfnamefont {P.~A.}\ \bibnamefont {Orellana}}, \ and\
  \bibinfo {author} {\bibfnamefont {J.~A.}\ \bibnamefont {Ot{\'a}lora}},\
  }\href@noop {} {\bibfield  {journal} {\bibinfo  {journal} {Annalen der
  Physik}\ }\textbf {\bibinfo {volume} {532}},\ \bibinfo {pages} {1900409}
  (\bibinfo {year} {2020})}\BibitemShut {NoStop}%
\bibitem [{\citenamefont {Wan}\ \emph {et~al.}(2005)\citenamefont {Wan},
  \citenamefont {Wei},\ and\ \citenamefont {Wang}}]{wan2005shot}%
  \BibitemOpen
  \bibfield  {author} {\bibinfo {author} {\bibfnamefont {L.}~\bibnamefont
  {Wan}}, \bibinfo {author} {\bibfnamefont {Y.}~\bibnamefont {Wei}}, \ and\
  \bibinfo {author} {\bibfnamefont {J.}~\bibnamefont {Wang}},\ }\href@noop {}
  {\bibfield  {journal} {\bibinfo  {journal} {Nanotechnology}\ }\textbf
  {\bibinfo {volume} {17}},\ \bibinfo {pages} {489} (\bibinfo {year}
  {2005})}\BibitemShut {NoStop}%
\bibitem [{\citenamefont {Flensberg}(2011)}]{PhysRevLett.106.090503}%
  \BibitemOpen
  \bibfield  {author} {\bibinfo {author} {\bibfnamefont {K.}~\bibnamefont
  {Flensberg}},\ }\href {\doibase 10.1103/PhysRevLett.106.090503} {\bibfield
  {journal} {\bibinfo  {journal} {Phys. Rev. Lett.}\ }\textbf {\bibinfo
  {volume} {106}},\ \bibinfo {pages} {090503} (\bibinfo {year}
  {2011})}\BibitemShut {NoStop}%
\bibitem [{\citenamefont {Liu}\ \emph {et~al.}(2015{\natexlab{b}})\citenamefont
  {Liu}, \citenamefont {Cheng},\ and\ \citenamefont
  {Lutchyn}}]{Lutchyn_Probing}%
  \BibitemOpen
  \bibfield  {author} {\bibinfo {author} {\bibfnamefont {D.~E.}\ \bibnamefont
  {Liu}}, \bibinfo {author} {\bibfnamefont {M.}~\bibnamefont {Cheng}}, \ and\
  \bibinfo {author} {\bibfnamefont {R.~M.}\ \bibnamefont {Lutchyn}},\ }\href
  {\doibase 10.1103/PhysRevB.91.081405} {\bibfield  {journal} {\bibinfo
  {journal} {Phys. Rev. B}\ }\textbf {\bibinfo {volume} {91}},\ \bibinfo
  {pages} {081405} (\bibinfo {year} {2015}{\natexlab{b}})}\BibitemShut
  {NoStop}%
\end{thebibliography}
\bibliographystyle{apsrev4-1}

\end{document}